\documentclass[12pt,a4paper]{article}              

\usepackage[left=25mm,right=25mm,top=3cm,bottom=3cm]{geometry}
\usepackage{graphicx}
\usepackage{subfig}
\usepackage{bm}
\usepackage{colortbl}
\usepackage{fancyhdr}
\usepackage{amssymb,amsfonts,amsmath,amsthm}
\usepackage{natbib}
\usepackage{bstnotations}
\usepackage{tabularx}

\usepackage{bbm}
\usepackage{booktabs} 
\usepackage{multirow} 
\usepackage{color}

\usepackage{authblk}

\graphicspath{{./},{./images/}}

\usepackage{url}
\usepackage{caption}

\newcommand{\uqlab}{\textsc{UQLab}}
\newcommand{\IN}{\textsc{input}}
\newcommand{\MO}{\textsc{model}}
\newcommand{\AN}{\textsc{analysis}}
\newcommand{\matlab}{\textsc{Matlab}}
\newcommand{\bfx}{\bm{x}}
\newcommand{\vX}{\ensuremath{\cx}}
\newcommand{\vY}{\ensuremath{\bm{y}}}
\newcommand{\bfth}{\bm{\theta}}
\renewcommand{\eqref}[1]{Eq.~(\ref{#1})}
\newcommand{\secref}[1]{Section~\ref{#1}}

\usepackage{fancyvrb}
\DefineVerbatimEnvironment{Code}{Verbatim}{}
\DefineVerbatimEnvironment{CodeInput}{Verbatim}{fontshape=sl}
\DefineVerbatimEnvironment{CodeOutput}{Verbatim}{}
\makeatletter
\newcommand\code{\bgroup\@makeother\_\@makeother\~\@makeother\$\@codex}
\def\@codex#1{{\normalfont\ttfamily\hyphenchar\font=-1 #1}\egroup}
\makeatother

\begin{document}

\title{The Gaussian process modelling module in UQLab}

\author{Christos Lataniotis}
\author{Stefano Marelli}
\author{Bruno Sudret}

\affil{Chair of Risk, Safety and Uncertainty Quantification,\\
    ETH Zurich, Stefano-Franscini-Platz 5, 8093 Zurich, Switzerland}

\date{}

\maketitle


\abstract{

We introduce the Gaussian process (GP) modelling module developed within the \uqlab{} software framework. The novel design of the GP-module aims at providing seamless integration of GP modelling into any uncertainty quantification workflow, as well as a standalone surrogate modelling tool. We first briefly present the key mathematical tools at the basis of GP modelling (a.k.a. Kriging), as well as the associated theoretical and computational framework. We then provide an extensive overview of the available features of the software and demonstrate its flexibility and user-friendliness. Finally, we showcase the usage and the performance of the software on several applications borrowed from different fields of engineering. These include a basic surrogate of a well-known analytical benchmark function, a hierarchical Kriging example applied to wind turbine aero-servo-elastic simulations and a more complex geotechnical example that requires a non-stationary, user-defined correlation function. The GP-module, like the rest of the scientific code that is shipped with \uqlab{}, is open source (BSD license).


}

\section{Introduction} \label{sec:Introduction}

Uncertainty quantification (UQ) through computer simulation is an interdisciplinary field that has seen a rapid growth in the last decades. Broadly speaking, it aims at i) identifying and quantifying the uncertainty in the input parameters of numerical models of physical systems, and ii) quantitatively assessing its effect on the model responses. Such a general formulation comprises a number of applications, including structural reliability \citep{Lemaire2009}, sensitivity analysis \citep{Saltelli2000}, reliability-based design optimisation \citep{Tsompanakis2008} and Bayesian techniques for calibration and validation of computer models \citep{Dashti2017}.

Due to the high cost of repeatedly evaluating complex computational models, analyses with classical sampling techniques such as Monte Carlo simulation are often intractable. In this context, meta-modelling techniques (also known as surrogate modelling) allow one to develop fast-to-evaluate surrogate models from a limited collection of runs of the original computational model, referred to as the experimental design \citep{santner_design_2003,Fang2005,Forrester2008}. Popular surrogate modelling techniques include Kriging \citep{Sacks1989}, polynomial chaos expansions \citep{Ghanembook1991,Xiu2002} and support vector regression \citep{Vapnik:1995}.

Kriging is a surrogate modelling technique first conceived by \citet{krige_1951} in the field of geostatistics and later introduced for the design and analysis of computer experiments by \citet{Sacks1989} and \citet{Welch1992}. The potential applications of Kriging in the context of civil and mechanical engineering, range from basic uncertainty propagation to reliability and sensitivity analysis \citep{marrel2008,Gaspar2014,Iooss2014,SudretHandbookUQ,MoustaphaJRUES2018}.
Beyond approximating the output of a computational model, Kriging surrogates also provide local estimates of their accuracy (via the variance of the Kriging predictor). This enables adaptive schemes \eg{}in the context of reliability analysis \citep{Echard2011,Dubourg2014} or surrogate model- based design optimisation \citep{Simpson2001,MalikiSMO2016}. The local error estimates of a Kriging surrogate have also led to improved Bayesian calibration of computer models (see \eg \citet{Bachoc2014a}). 

Although in its standard form Kriging is a stochastic interpolation method, certain extensions have been proposed for dealing with noisy observations. Such extensions have been of particular interest to the machine learning community and they are commonly referred to as \emph{Gaussian process regression} \citep{Rasmussen2006}.

 A number of dedicated toolboxes are readily available for calculating Kriging surrogate models. Of interest to this review is general purpose software not targeted to specific Kriging applications, because they are typically limited to two or three dimensional problems (see \eg  gslib \citep{deutsch1992gslib}). 
  Within the \textsc{R} community one of the most comprehensive and well-established Kriging packages is arguably DiceKriging, developed by the DICE consortium \citep{roustant_dicekriging_2012}.  This set of packages provides Kriging meta-modelling as part of a framework for adaptive experimental designs and Kriging-based optimisation based on the packages DiceDesign and DiceOptim \citep{rDiceDesign,rDiceOptim}. 
\textsc{scikit-learn} provides a \textsc{python}-based, machine-learning-oriented implementation of Gaussian processes for regression and classification \citep{scikit-learn}. Alternatively, PyKriging \citep{paulson_pykriging:_2015} offers a Kriging toolbox in \textsc{python} that offers basic functionality with focus on user-friendliness. Gpy \citep{gpy_2012} offers a Gaussian process framework with focus on regression and classification problems. 
Within the \textsc{Matlab} programming language the first Kriging toolbox with widespread use was DACE \citep{lophaven_aspects_2002}. DACE was later extended to ooDACE \citep{couckuyt_oodace_2014}, an object-oriented Kriging implementation with a richer feature set. Small Toolbox for Kriging \citep{STK} offers an alternative Kriging implementation that is mainly focused on providing a set of functions for Kriging surrogate modelling and design of experiments. GPML \citep{rasmussen_gaussian_2010} offers a library of functions that are directed towards Gaussian processes for regression and classification in a machine learning context. Finally, recent versions of \matlab{} (starting from R2015b) provide a rapidly growing Gaussian process library for regression and classification.

Due to the variety of potential applications of Kriging, different toolboxes tend to be focused on a specific user niche. There is limited availability of general purpose Kriging toolboxes that allow for seamless integration within various UQ workflows ranging from \eg basic uncertainty propagation to reliability analysis and surrogate-model-based optimisation. To this end, the Kriging toolbox presented here was developed as a module of the general purpose UQ framework, \uqlab{} (\cite{Marelli2014}, \url{www.uqlab.com}). In addition, although most of the aforementioned toolboxes offer a significant set of configuration options, the support for fully customisable Kriging is often limited or not easily accessible, which can be a drawback in a research environment. Finally, the user experience may vary from user-friendly to complex (especially to access the most advanced features), often requiring a significant degree of programming knowledge. This might be rather inconvenient for applied scientists and practitioners with limited programming knowledge. 
Following these premises, this paper introduces the \uqlab{} Gaussian process modelling tool (GP-module) focusing on its unique embedding into a complex uncertainty quantification environment, its user-friendliness and customisability. 

The paper is structured as follows: in \secref{sec:Kriging} a theoretical introduction to Kriging is given to highlight its main building blocks. 
In  \secref{sec:UQLabKrigingHead} the key-features of the GP-module are presented. Finally, a set of application examples is used to showcase in detail the usage of the software in  \secref{sec:Applications}, followed by a summary and a road map of the upcoming developments in \secref{sec:Summary}.

\section{Kriging theory} \label{sec:Kriging}

\subsection{Kriging basics} \label{sec:basic_equations}
Any metamodeling approach, such as Kriging, aims at approximating the response of a computational model given a finite set of observations. In this context, consider a system whose behaviour is represented by a computational model $\cm$ which maps the $M$-dimensional input parameter space $\cd_{\ve{x}}$ to the $1$-dimensional output space, \ie $\cm \, : \, \bfx \in \cd_{\ve{x}} \subset \mathbb{R}^M \mapsto y \in \mathbb{R}$ where $\bfx = \left\lbrace  x_1 \enu x_M \right\rbrace^\top$.

\begin{figure}[!ht]
  \captionsetup{width=14cm}
  \centering
	\includegraphics[height=6.5cm,keepaspectratio]{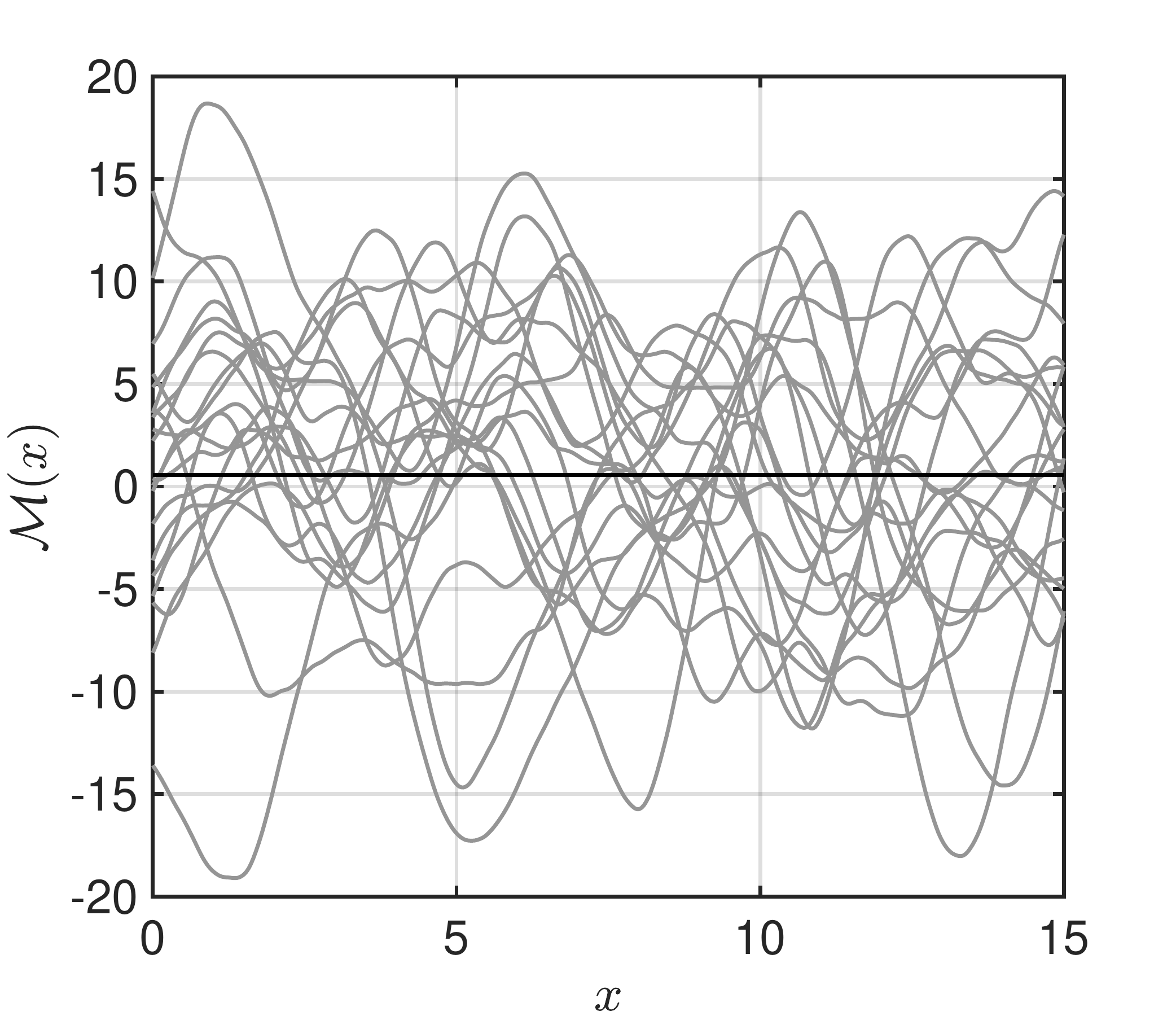}
	\includegraphics[height=6.5cm,keepaspectratio]{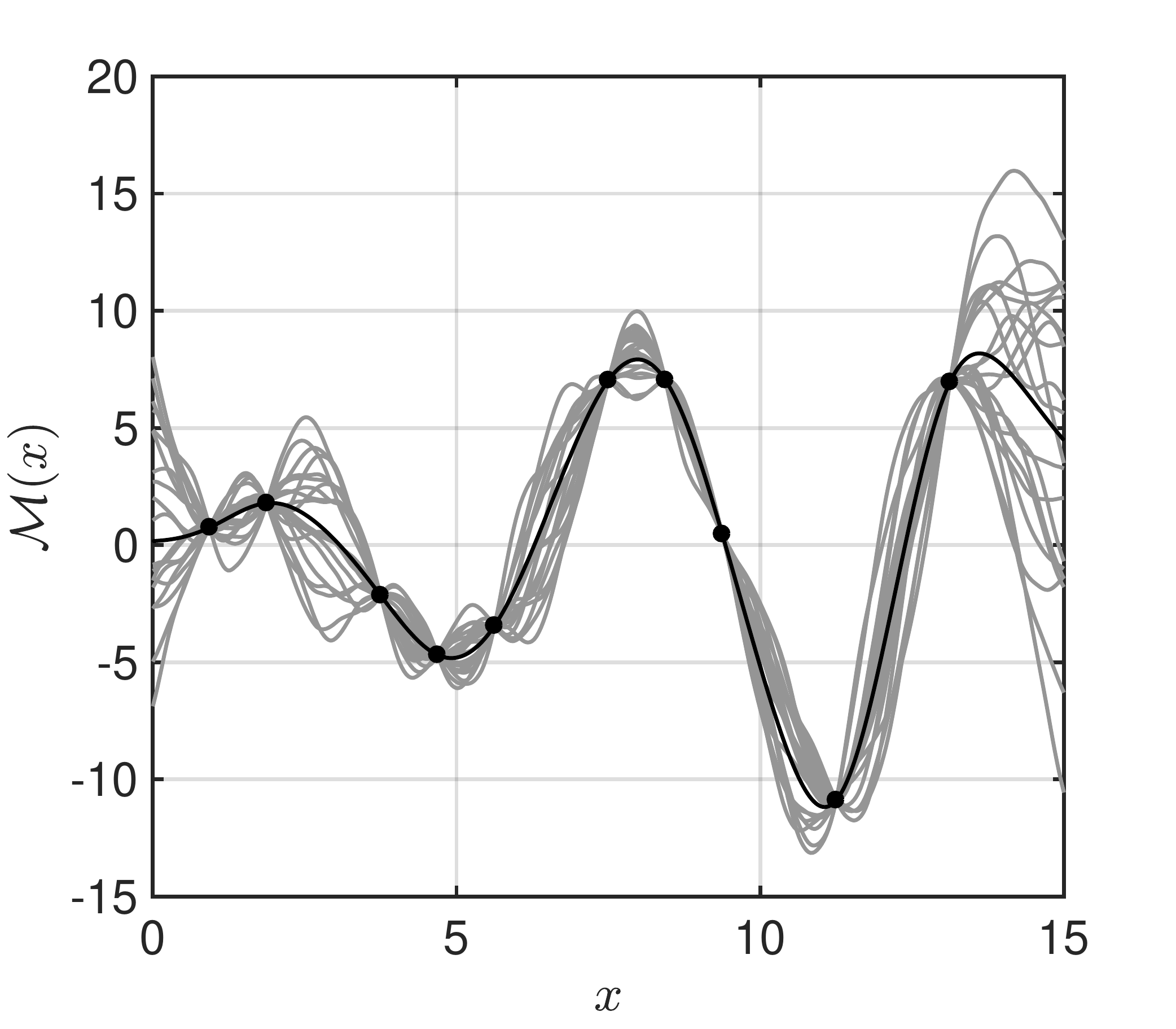}
	\caption{Realisations of a prior (left) and posterior Gaussian process (right). The Gaussian process mean in each case is denoted by a black line.}
	\label{fig:krg_example}
\end{figure}

Kriging is a meta-modelling technique which assumes that the true model response is a realisation of a Gaussian process described by the following equation \citep{santner_design_2003}:

\begin{equation} \label{eq:KrigingGeneral}
	\cm^K(\bfx) = \ve{\beta}^\top \ve{f}(\ve{x}) + \sigma^2 Z(\bfx,\omega)
\end{equation}

where $\ve{\beta}^\top \ve{f}(\ve{x})$ is the mean value of the Gaussian process, also called \emph{trend},  $\sigma^2$ is the Gaussian process variance  and $Z(\bfx, \omega)$ is a zero-mean, unit-variance Gaussian process. This process is fully characterised by the auto-correlation function between two sample points $R(\bfx,\bfx';\bfth)$. The hyperparameters $\bfth$ associated with the correlation function $R(\cdot;\bfth)$ are typically unknown and need to be estimated from the available observations. 

Having specified the trend and the correlation function parameters it is possible to obtain an arbitrary number of realisations of the so-called \emph{prior} Gaussian process (see Figure~\ref{fig:krg_example} left). In the context of metamodelling the goal is to calculate a prediction $\cm^{K}(x)$ for a new point $\bfx$, given an experimental design $\vX = \acc{\bfx^{(1)} \enu \bfx^{(N)} }$ of size $N$ and  the corresponding (noise-free) model responses $\vY = \{y^{(1)}=\cm(\bfx^{(1)}) \enu   y^{(N)}= \cm(\bfx^{(N)})\}^\top$. A Kriging metamodel (a.k.a. Kriging predictor) provides such predictions based on the properties of the so-called \emph{posterior} Gaussian process conditioned on the available data (see Figure~\ref{fig:krg_example} right). The Kriging prediction on $\bfx$ corresponds to a random variate $\widehat{Y}(\bfx) \sim \cn \left( \mu_{\widehat{Y}}(\ve{x}), \sigma_{\widehat{Y}}(\ve{x}) \right)$, therefore the approximation of the computational model that is obtained is essentially an infinite family of such models. Each of these models is a realisation (or sample) of the posterior Gaussian process. In practice the mean response is used (see  \eqref{eq:TheoryPredicorMean}) as the Kriging surrogate, while its variance (see  \eqref{eq:TheoryPredicorVariance}) is often interpreted as a measure of the local error of the prediction. The equations for calculating the mean and variance of a universal Kriging predictor are given next.

The Gaussian assumption states that the vector formed by the true model responses, $\vY$ and the prediction, $\widehat{Y}(\bfx)$, has a joint Gaussian distribution defined by:

\begin{equation} 
\left\lbrace  \begin{matrix}
\widehat{Y}(\bfx) \\ \vY
\end{matrix}\right\rbrace 
\sim \mathcal{N}_{N+1} \left (  
\left\lbrace  \begin{matrix}
\bm{f}^\top(\bfx) \bm{\beta} \\ \mat{F} \bm{\beta}
\end{matrix}\right\rbrace, 
\sigma^2 \left\lbrace 
\begin{matrix}
1 & \bm{r}^\top(\bfx) \\ 
\bm{r}(\bfx) & \mat{R}
\end{matrix}
\right\rbrace
\right )
\end{equation}

where $\bm{F}$ is the information matrix of generic terms:
\begin{equation}
	F_{ij}  = f_j(\ve{x}^{(i)})~,~i=1 \enu N,~j=1 \enu P,
\end{equation}

$\bm{r}(\bfx)$ is the vector of cross-correlations between the prediction point $\bfx$ and each one of the observations whose terms read:
\begin{equation}
	r_{i}(\bfx) = R(\bfx,\bfx^{(i)};\bm{\theta}), ~i=1 \enu N.
	\label{eq:r0}
\end{equation}

$\bm{R}$ is the correlation matrix given by:
\begin{equation}
	R_{ij} = R(\bfx^{(i)},\bfx^{(j)};\bm{\theta}), ~i,j=1 \enu N.
\end{equation}

The mean and variance of the Gaussian random variate $\widehat{Y}(\bfx)$ (a.k.a. mean and variance of the Kriging predictor) can be calculated based on the best linear unbiased predictor properties \citep{santner_design_2003}:
\begin{equation} \label{eq:TheoryPredicorMean}
\mu_{\widehat{Y}}(\ve{x})  =\ve{f}(\ve{x})^\top \ve{\beta} + \ve{r}(\ve{x})^\top \mat{R}^{-1}\left (\vY-\mat{F}\ve{\beta} \right )\, ,
\end{equation}
\begin{equation} \label{eq:TheoryPredicorVariance}
\sigma_{\widehat{Y}}^2(\ve{x})  = \sigma^2 \left(  1-\ve{r}^\top(\ve{x})\mat{R}^{-1}\ve{r}(\ve{x}) + \ve{u}^\top(\ve{x}) (\mat{F}^\top\mat{R}^{-1}\mat{F})^{-1}\ve{u}(\ve{x})  \right)
\end{equation}
where:
\begin{equation} \label{eq:AKG:TheoryCalcBeta}
\ve{\beta}  = \left( \mat{F}^\top \mat{R}^{-1} \mat{F}  \right)^{-1}\mat{F}^\top\mat{R}^{-1} \vY
\end{equation}
is the generalised least-squares estimate of the underlying regression problem and 
\begin{equation}
\label{eq:AKG:TheoryPredicorU}
\ve{u}(\ve{x}) = \mat{F}^\top \mat{R}^{-1}\ve{r}(\ve{x}) - \ve{f}(\ve{x}).
\end{equation}
 
Once $\mu_{\widehat{Y}}(\ve{x})v$ and $\sigma_{\widehat{Y}}^2(\ve{x})$ are available, confidence bounds on predictions can be derived by observing that:

\begin{equation}
	\label{eq:gaussianity}
\cp\left[\widehat{Y}(\ve{x}) \leq t \right] = \Phi \left( \frac{t-\mu_{\widehat{Y}}(\ve{x})}{\sigma_{\widehat{Y}}(\ve{x})}\right) ,
\end{equation}
where $\Phi(\cdot)$ denotes the Gaussian cumulative distribution function. Based on \eqref{eq:gaussianity} the confidence intervals on the predictor can be calculated by:
\begin{equation}\label{eq:AKG:TheoryConfIntervals}
\widehat{Y}(\ve{x}) \in \left[ \mu_{\widehat{Y}}(\ve{x}) - \Phi^{-1}\left( 1 - \frac{\alpha}{2} \right)\sigma_{\widehat{Y}}(\ve{x}) ~,~ 
\mu_{\widehat{Y}}(\ve{x}) + \Phi^{-1}\left( 1 - \frac{\alpha}{2} \right)\sigma_{\widehat{Y}}(\ve{x}) \right]
\end{equation}
and can be interpreted as the interval within which the Kriging prediction falls with probability $1-\alpha$.

The equations that were derived for the best linear unbiased Kriging predictor  assumed that the covariance function $\sigma^2 R(\cdot;\bfth)$ is known. In practice however, the family and other properties of the correlation function need to be selected \textit{a priori}. The hyperparameters $\bm{\theta}$, the regression coefficients $\bm{\beta}$  and the variance $\sigma^2$ need to be estimated based on the available experimental design. This involves solving an optimisation problem that is further discussed in \secref{Ingredients_estim}. The resulting best linear unbiased predictors are called \emph{empirical} in \citet{santner_design_2003} because they typically result from empirical choice of various Kriging parameters that are further discussed in Sections \ref{Ingredients_trend} - \ref{Ingredients_estim}.

\subsection{Trend} \label{Ingredients_trend}

The trend refers to the mean of the Gaussian process, \ie the $\ve{\beta}^\top \ve{f}(\ve{x})$  term in \eqref{eq:KrigingGeneral}. Using a non-zero trend is optional but it is often preferred in practice (see \eg \citet{Rasmussen2006,SchoebiIJUQ2015}). Note that the mean of the Kriging predictor in \eqref{eq:TheoryPredicorMean} is not confined to be zero when the trend is zero.  

In the literature, it is customary to distinguish between Kriging metamodels depending on the type of trend they use \citep{stein_interpolation_1999,santner_design_2003,Rasmussen2006}.
The most general and flexible formulation is \emph{universal Kriging}, which assumes that the trend is composed of a sum of $P$ arbitrary functions $f_k(\bfx)$, \ie
\begin{equation} \label{trend:general_formula}
	\bm{\beta}^\top \bm{f}(\bfx) = \sum_{k=1}^{P} \beta_k f_k(\bfx).
\end{equation}
Some of the most commonly used trends for universal Kriging are given for reference in \tabref{tbl:TheoryTrendTypes}. 
\emph{Simple Kriging} assumes that the trend has a known constant value, \ie  $P=1$, $f_1(\bfx)=1$ and $\beta_1$ is known. In \emph{Ordinary Kriging} the trend has a constant but unknown value, \ie $P=1$, $f_1(\bfx)=1$ and $\beta_1$ is unknown.

\begin{table}[!ht]
\centering
\begin{tabular}{l l}
	\hline

Trend & Formula  \\ \hline
constant (ordinary Kriging)&
$\beta_0$ \\
	
linear &
$\beta_0 + \sum\limits_{i=1}^{M}\beta_i x_i$  \\
	
quadratic &
$\beta_0 + \sum\limits_{i=1}^{M}\beta_i x_i+ \Sigma_{i=1}^{M}\Sigma_{j=1}^{M}\beta_{ij} x_i x_j$ \\

\hline
\end{tabular}

\caption{Formulas of the most commonly used Kriging trends.}
\label{tbl:TheoryTrendTypes}

\end{table}

\subsection{Correlation function}\label{Ingredients_corr}

\begin{figure}[!ht]  \captionsetup{width=14cm}
	\centering
	\includegraphics[height=6.5cm,keepaspectratio]{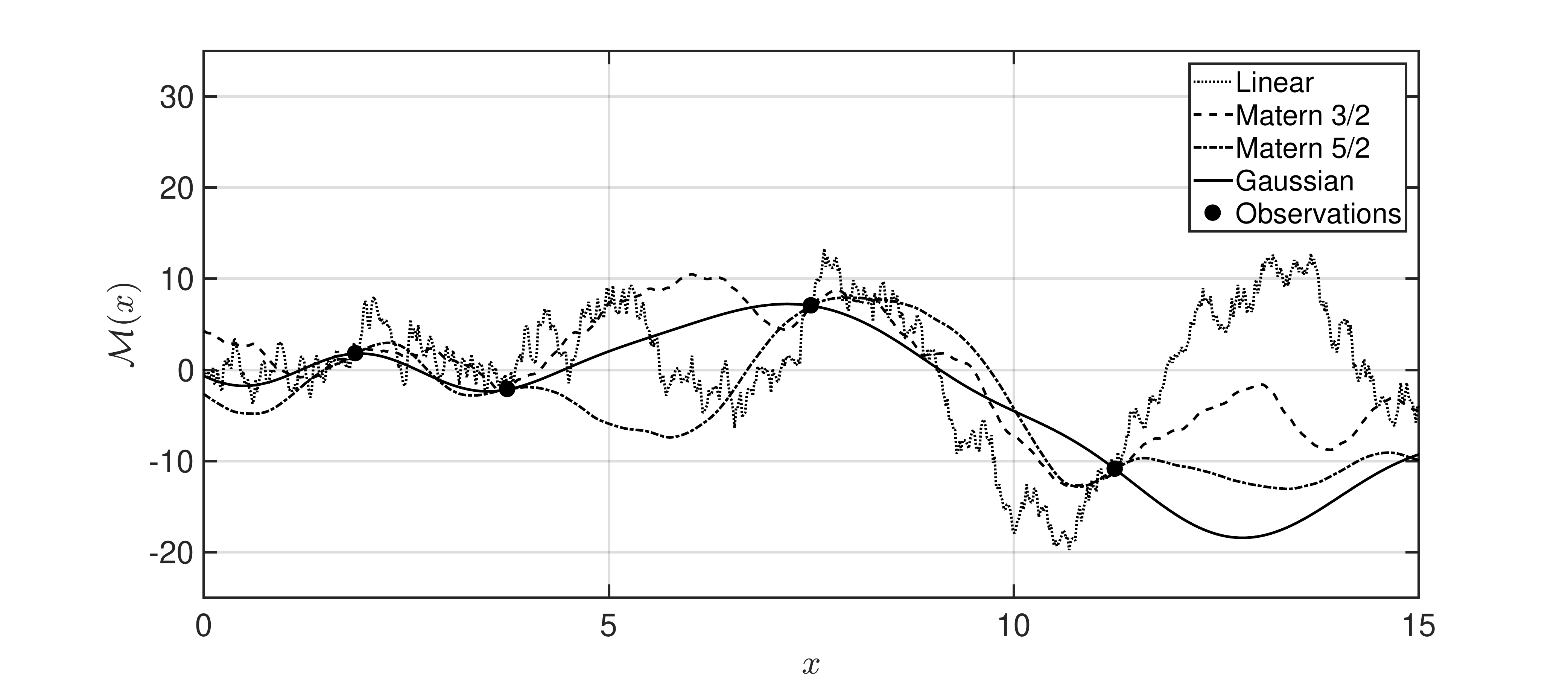}
	\caption{Realisations of Gaussian processes, characterised by various correlation families and the same length-scale ($\theta$) value.}
	\label{fig:gp_smoothness}
\end{figure}

The correlation function (also called \emph{kernel} in the literature, or \emph{covariance} function if it includes the Gaussian process variance $\sigma^2$) is a crucial ingredient for a Kriging metamodel, since it contains the assumptions about the function that is being approximated. An arbitrary function of $(\ve{x},\ve{x}')$ is in general not a valid correlation function. In order to be admissible, it has to be chosen in the set of positive definite kernels. However, checking for positive definiteness of a kernel can be a challenging task. Therefore it is usually the case in practice to select families of kernels known to be positive definite and to estimate their parameters based on the available experimental design and model responses (see \secref{Ingredients_estim}). A usual assumption is to consider kernels that depend only on the quantity $h = \norme{\ve{x} - \ve{x}'}{}$ which are called \emph{stationary}. A list of stationary kernels commonly used in the literature can be found in \tabref{tbl:TheoryCorrFamilies}.  Different correlation families result in different levels of smoothness for the associated Gaussian processes, as depicted in \figref{fig:gp_smoothness} \citep{Rasmussen2006}.

\begin{table}[!ht]
\centering
\begin{tabular}[t]{l l}
	\hline
	
	Name & Formula  \\ \hline
	Linear &
	$R(h;\theta) = \max \left(  0, 1 -    \frac{\left| h \right|}{\theta}    \right)$ \\
	
	Exponential&
	$R(h;\theta) = \exp\left(  -   \frac{\left| h \right|}{\theta}    \right)$ \\
	
	Mat\'ern $3/2$ &
	$R(h;\theta) = \left(1+ \frac{\sqrt{3}|h|}{\theta} \right) \exp\left( -\frac{\sqrt{3}|h|}{\theta} \right)$ \\
	
	Mat\'ern $5/2$ &
	$R(h;\theta) = \left(1+ \frac{\sqrt{5}|h|}{\theta} + \frac{5 h^2}{3\theta^2} \right)
	\exp\left( -\frac{\sqrt{5}|h|}{\theta} \right)$ \\
	
	Gaussian (squared exponential)&
	$R(h;\theta) = \exp\left(  - \sum_{i=1}^{M}  \left( \frac{ h }{\theta}   \right)^2   \right)$ \\
	
	\hline
\end{tabular}
\caption{List of available correlation families.}
\label{tbl:TheoryCorrFamilies}
\end{table}

In case of multidimensional inputs ($M>1$) it is common practice to obtain admissible kernels as functions of one-dimensional correlation families as the ones in \tabref{tbl:TheoryCorrFamilies}. Two standard approaches in the literature are the \emph{separable} correlation type \citep{Sacks1989}:
\begin{equation}\label{eq:corr_separable}
	R(\ve{x},\ve{x}';\ve{\theta}) = \prod\limits_{i=1}^{M} R(x_i,x_i',\theta_i)
\end{equation}
  and the \emph{ellipsoidal} type \citep{Rasmussen2006}:
\begin{equation}\label{eq:corr_ellipsoidal}
  	R(\ve{x},\ve{x}';\ve{\theta}) = R(h)~,~h = \sqrt{\sum\limits_{i=1}^{M}
		\left(\frac{x_i- x_i'}{\theta_i} \right)^2~}.
\end{equation} 

Although typically $\ve{\theta} \in \mathbb{R}^M$ this is not necessarily true in the general case, since the number of components of $\bm{\theta}$ that correspond to each input dimension may vary. In the current stage, it is assumed however that one element of $\ve{\theta}$ is used per dimension for notational clarity. 

In certain scenarios (\eg based on prior knowledge), \emph{isotropic} correlation functions can be used for multidimensional inputs. In that case the same correlation function parameters $\bm{\theta}$ are used for each input dimension in \eqref{eq:corr_separable} and \eqref{eq:corr_ellipsoidal}.

\subsection{Estimating the hyperparameters} \label{Ingredients_estim}

In most practical applications of Kriging surrogate modelling, the hyperparameters $\ve{\theta}$ are estimated given an experimental design $\cx$ and model responses $\ve{y}$.  \textit{Maximum likelihood} and \textit{cross-validation} are the most commonly used methods for doing so and further discussed next. 

The \emph{maximum likelihood} approach aims at finding the set of parameters $\ve{\beta}, \ve{\theta}, \sigma^2$ such that the likelihood of the observations $\ve{y} = \left \lbrace \cm(\bfx_1) \enu \cm(\bfx_N)  \right\rbrace^\top$ is maximal. Since $\ve{y}$ follows a multivariate Gaussian distribution, the likelihood function reads:

\begin{equation} \label{eq:Likelihood}
L(\vY ~|~\ve{\beta}, \sigma^2 , \ve{\theta}) = \frac{\text{det}(\mat{R})^{-1/2}}{(2\pi\sigma^2)^{N/2} } \,\text{exp} \left[-\frac{1}{2\sigma^2}(\vY-\mat{F} \ve{\beta})^\top \mat{R} ^{-1}(\vY-\mat{F} \ve{\beta})  \right].
\end{equation}

For any given value of $\ve{\theta}$, the maximisation of the likelihood \emph{w.r.t.} $\ve{\beta}$ and $\sigma^2$ is a convex quadratic programming problem. Consequently, it admits closed form generalized least-squares estimates of $\ve{\beta}$ and $\sigma^2$ (for proof and more details see \eg  \citet{santner_design_2003}):

\begin{equation} \label{eq:betaML}
\ve{\beta} = \ve{\beta}(\ve{\theta})  = \left( \mat{F}^\top \mat{R}^{-1}  \mat{F}  \right)^{-1}\mat{F}^\top\mat{R}^{-1}  \vY \, ,
\end{equation} 
\begin{equation} \label{eq:sigma2ML}
\sigma^2 = \sigma^2(\ve{\theta}) = \frac{1}{N}  \left( \vY-\mat{F} \ve{\beta} \right)^\top \mat{R} ^{-1} \left( \vY-\mat{F} \ve{\beta}  \right).
\end{equation} 

The value of the hyperparameters  $\ve{\theta}$ is calculated by solving the optimisation problem:

\begin{equation} \label{eq:thetaML_short}
\ve{\theta} =\arg~\underset{\cd_{\ve{\theta}}}{\min} ~ \left(  -\log L(\vY ~|~\ve{\beta}, \sigma^2 , \ve{\theta}) \right).
\end{equation}

Based on Eqs~(\ref{eq:Likelihood}) - (\ref{eq:sigma2ML}) the optimisation problem in \eqref{eq:thetaML_short} can be written as follows:

\begin{equation} \label{eq:thetaML}
\ve{\theta} = \arg~\underset{\cd_{\ve{\theta}}}{\min} ~\left(\frac{1}{2}\log(\text{det}(\mat{R})) + \frac{N}{2}\log(2\pi\sigma^2)+\frac{N}{2}  \right).
\end{equation}

\vspace{.3cm}

The \emph{cross-validation} method (also known as $K$-fold cross-validation) is based instead on partitioning the whole set of observations $\cs \eqdef \acc{\cx, \ve{y}}$ into $K$ mutually exclusive and collectively exhaustive subsets $\left\lbrace  \cs_k, k=1,\ldots,K   \right\rbrace$ such that 
\begin{equation} \label{eq:CVgeneral}
\cs_i \cap  \cs_j = \emptyset ~, ~ \forall (i,j) \in \lbrace 1,\ldots,K \rbrace ^2 ~ \text{ and } \bigcup_{k=1}^{K} \cs_k = \cs .
\end{equation}

The $k$-th set of cross-validated predictions is obtained by calculating the Kriging predictor using all the subsets but the $k$-th one and evaluating its predictions on that specific $k$-th fold that was left apart. The leave-one-out cross-validation procedure corresponds to the special case that the number of classes is equal to the number of observations ($K = N$).

In the latter case the objective function is \citep{santner_design_2003,Bachoc2013b}:

\begin{equation} \label{eq:TheoryCV_J}
\ve{\theta} = \arg~\underset{\cd_{\ve{\theta}}}{\min} \sum_{i=1}^{K} \left( \cm(\ve{x}^{(i)}) - \mu_{\widehat{Y}, (-i)}(\ve{x}^{(i)})  \right) ^2   
\end{equation}

where $\mu_{\widehat{Y}, (-i)}(\ve{x}^{(i)})$ is the mean Kriging predictor that was calculated using $\cs \setminus \acc{\ve{x}^{(i)} , \ve{y}^{(i)}}$ evaluated at point $\ve{x}^{(i)}$. Notice that for the case of leave-one-out cross-validation, $i$ is an index but in the general case $i$ is a vector of indices. Calculating the objective function in \eqref{eq:TheoryCV_J} requires the calculation of $K$ Kriging surrogates. The computational requirements for performing this operation can be significantly reduced as shown in \citet{Dubrule1983}.  


The estimate of $\sigma^2$ is calculated using the following equation \citep{ cressie_statistics_1993,Bachoc2013b}:

\begin{equation} \label{eq:CV_sigma2}
\sigma^2 = \sigma^2(\ve{\theta}) = \frac{1}{K} \sum_{i=1}^{K} \frac
{ \left( \cm(\ve{x}^{(i)}) - \mu_{\widehat{Y}, (-i)}(\ve{x}^{(i)})  \right) ^2}
{\sigma^2_{\widehat{Y}, (-i)}(\ve{x}^{(i)})}
\end{equation}

where $\sigma^2_{\widehat{Y}, (-i)}(\ve{x}^{(i)})$  denotes the variance of a Kriging predictor that was calculated using $\cs \setminus \acc{\ve{x}^{(i)} , \ve{y}^{(i)}}$, evaluated at point $\ve{x}^{(i)}$. When $i$ is a set of indices, the division and the squared operations in \eqref{eq:CV_sigma2} are performed element-wise.

Numerically solving the optimisation problems described in \eqref{eq:thetaML} (maximum likelihood case) or \eqref{eq:TheoryCV_J} (cross-validation case) relies on either local (\eg gradient-based) or global (\eg evolutionary) algorithms. On the one hand, local methods tend to converge faster and require fewer objective function evaluations than their global counterparts. On the other hand, the existence of flat regions and multiple local minima, especially for larger input dimension, can lead gradient methods to poor performance when compared to global methods.  It is common practice to combine both strategies sequentially to improve global optimisation results with a final local search (which is also known as \emph{hybrid} methods).

It can be often the case in engineering applications that different components of the input variable $\bfx$ take values that differ by  orders of magnitude. In such cases, potential numerical instabilities can be avoided by scaling $\cx \mapsto \cu$, \eg as follows:

\begin{equation}\label{eq:scalingX}
 u^{(i)}_j = \frac{x^{(i)}_j - \Esp{\cx_j}}{\sqrt{\Var{\cx_j}}} \, , \, i= 1 \enu N\, , \, j= 1 \enu M  
\end{equation} 

where $u^{(i)}_j$ (resp. $x^{(i)}_j$) refer to the $i$-th sample of the $j$-th component of $\cu$ (resp. of $\cx$) and $\Esp{\cx_j}$ and $\Var{\cx_j}$ refer to the empirical mean and variance of the $j$-th component of $\cx$.

\section{The UQLab Gaussian process modelling module} \label{sec:UQLabKrigingHead}

\subsection{The UQLab project}

\begin{figure}[!ht]    
  \captionsetup{width=14cm}
  \centering 
  \subfloat[The theoretical UQ framework based on which any UQ problem can be described.]
  {\label{fig:uq_framework} \includegraphics[width=14cm,trim={0 15cm 2cm 0},clip]{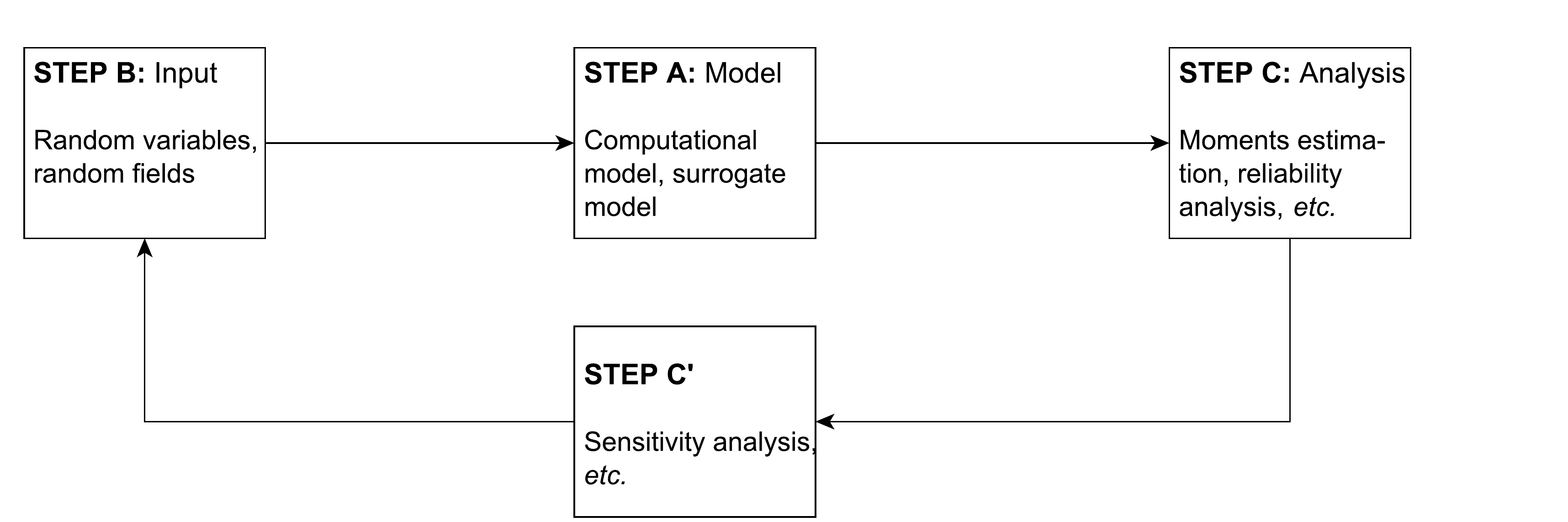}}

  \subfloat[The modular structure of the \uqlab{} framework. An
  arbitrary number of objects (Input, Model, Analysis) can be connected
  at any stage of the UQ problem.] 
  {\label{fig:uqlab_corescheme}
    \includegraphics[width=14cm,trim={0 0 0  13cm},clip]{uqlab_scheme_V2.pdf}}
%
%
%
	\caption{An abstract illustration of the \uqlab{} architecture (b) based on the theoretical UQ framework in (a) by \citet{SudretHDR}.}
	\label{fig:UQscheme }
\end{figure}

\uqlab{} is a software framework developed by the Chair of Risk, Safety and Uncertainty Quantification at ETH Z\"urich \citep{Marelli2014}. The goal of this project is to provide an uncertainty quantification tool that is accessible also to a non-highly-IT trained scientific audience. Due to the broadness of the UQ scope, a correspondingly general theoretical framework is required. The theoretical backbone of the \uqlab{} software lies in the global uncertainty framework developed by \citet{SudretHDR,Derocquigny2008}, sketched in \figref{fig:uq_framework}. According to this framework, the solution of any UQ problem can generally be decomposed into the following steps:

\vspace{0.4cm}

\hspace{-0.5cm}
\begin{tabularx}{1.0\linewidth}{l @{\extracolsep{0.4cm}} X}
\bf{Step A}&
Define the \emph{physical model} and the quantities of interest for the analysis. It is a deterministic representation of an arbitrarily complex physical model, \eg a finite element model in civil and mechanical engineering. In this category also lie metamodels, such as Kriging, since once they are calculated they can be used as surrogates of the underlying ``true'' model.    
\\

\bf{Step B}&
Identify and quantify the sources of uncertainty in the parameters of the system that serve as input for Step A. They are represented by a set of random variables and their joint probability density function (PDF).
\\
\bf{Step C}&
Propagate the uncertainties identified in Step B through the computational model in Step A to characterise the uncertainty in the model response. This type of analyses include moments analysis, full PDF characterisation, rare events estimation, sensitivity analysis, etc.
\\
\bf{Step C'}&
Optionally, exploit the by-products of the analysis in Step C to update the sources of uncertainty, \eg{} by performing model reduction based on sensitivity analysis.
\end{tabularx}

\vspace{0.4cm}

These components introduce a clear semantic distinction between the elements involved in any UQ problem: \MO{}, \IN{} and \AN{}. This theoretical framework provides the ideal foundation for the development of the information flow model in a multi-purpose UQ software. 

At the core of \uqlab{} lies a modular infrastructure that closely follows the semantics previously described, graphically represented in \figref{fig:uqlab_corescheme}. The three steps identified in \figref{fig:uq_framework} are directly mapped to \emph{core modules} in \figref{fig:uqlab_corescheme}: \textsc{model} corresponds to Step A (physical modelling, metamodeling), \textsc{input} to Step B (sources of uncertainty) and  \textsc{analysis} to Step C (uncertainty analysis). Within the \uqlab{} framework, a \emph{module} refers to some particular functionality, \eg the GP-module provides Kriging surrogate modelling. Each module extends the functionalities of one of the core modules. It can be either self-contained or capitalise on other modules for extended functionalities. 

The real ``actors'' of a UQ problem are contained in the \emph{objects} connected to each of the core modules. A typical example of such objects would be an \IN{} object that generates samples distributed according to arbitrary PDFs, a \MO{} object that runs a complex FEM simulation, or an \AN{} object that performs reliability analysis. The platform allows one to define an arbitrary number of objects and select the desired ones at various stages of the solution of a complex UQ problem. 

\uqlab{} first became freely available to the academic community on July 2015 as a  beta version. On April 2017  the version 1.0 of \uqlab{} was released. Starting from version 1.0 all the scientific code of the software is open-source (BSD license). By May 2018 around $1300$ users have already registered and used it.

\subsection{The GP-module} \label{sec:UQLabKriging}

Kriging is one of the metamodelling modules available in \uqlab{} \citep{UQdoc_10_105}. Following the semantics described in the previous section, it is attached to the \MO{} core module. Although the GP-module itself can be used by other modules, \eg an \AN{} module performing  reliability analysis combining Kriging and Monte Carlo Simulation (AK-MCS) \citep{Echard2011,UQdoc_10_107}, the focus of this work is on the capabilities of the GP-module itself. 

An overview of the available features of the GP-module is given in \tabref{tbl:UQKrigingFeatures}. The GP-module incorporates the four ingredients identified in \secref{sec:basic_equations}:
\begin{itemize}
    \item \textit{Trends}: Universal Kriging trends are fully supported, including simple, ordinary, or polynomial of arbitrary degree. In addition, custom basis functions $f(\bfx)$ or a completely custom trend function may be specified
    \item \textit{Correlation functions}: Standard correlation families from the literature are readily available as well as the possibility of creating user-defined ones. For multi-dimensional inputs ellipsoidal and separable correlation functions can be used, allowing also for isotropic ones. Fully user-specified correlation functions are also supported
    \item \textit{Estimation methods}: Maximum likelihood (\eqref{eq:thetaML}) and cross-validation (\eqref{eq:TheoryCV_J}) methods can be used for estimating the hyper-parameters
    \item \textit{Optimisation methods}: \matlab{}'s built-in local and global optimisation methods are offered, namely BFGS and genetic algorithm as well as genetic algorithm with BFGS refinement (hybrid).
\end{itemize}
In addition, various scaling operations are allowed for avoiding numerical instabilities during the hyperparameters estimation. Such operations may vary from simple zero-mean scaling to more advanced ones such as isoprobabilistic transformations by interfacing with other \uqlab{} modules.

Following the general design principle of \uqlab{} concerning user-friendliness, all the possible configuration options have default values pre-assigned to allow basic usage of the module with very few lines of code (see \secref{sec:Ex1:Basic}). A \matlab{} structure variable is used to specify a Kriging configuration, called \code{KOptions} in the following sections. 

To showcase the minimal working code for obtaining a Kriging surrogate a simple application is considered. The experimental design consists of $8$ random samples in the $[0,15]$ interval and it is contained in the variable \code{XED}. The ``true'' model is $\cm(x) = x \sin(x)$ and the corresponding model responses are stored in the variable \code{YED}. The minimal code required for obtaining a Kriging surrogate, given \code{XED} and \code{YED}  is the following:

\begin{table}[!ht]
\centering
\begin{tabularx}{\textwidth}{lllX}
\hline
\textbf{Feature} &\textbf{Specification}  & \textbf{Value}  & \textbf{Description}  \\
\hline
Trend & & Simple   &  A constant term specified by the user (simple Kriging)\\
      & & \textbf{Ordinary} &  A constant term estimated using \eqref{eq:AKG:TheoryCalcBeta} (ordinary Kriging) \\
      & & Polynomial basis  & The trend in \eqref{trend:general_formula} consists of polynomial basis functions $f_k$ of arbitrary degree  \\
      & & Custom basis   &  The trend in \eqref{trend:general_formula} consists of arbitrary functions $f_k$ \\
      & & Custom trend   &  Custom trend function that computes $\bm{F}$ directly\\
\hline
Correlation &  Types    & Separable  & As described in \eqref{eq:corr_separable}. Both isotropic and anisotropic     variants are supported. \\
            &           & \textbf{Ellipsoidal}  & As described in \eqref{eq:corr_ellipsoidal}. Both isotropic and anisotropic variants are supported. \\
            &           & Custom  &  Custom correlation function that computes $\bm{R}$ directly \\
            &  Families & Commonly used & All the correlation families reported in \tabref{tbl:TheoryCorrFamilies} are available \\
            &           & Custom & A custom correlation family can be specified  \\
\hline
Estimation &      &  ML & Maximum-likelihood estimation (see \eqref{eq:thetaML}) \\
           &      &  \textbf{CV} & $K$-fold Cross-Validation method (see \eqref{eq:TheoryCV_J}). Any $K$ value is supported \\
\hline
Optimisation &      & BFGS  & Gradient-based optimisation method (Broyden-Fletcher-Goldfarb-Shanno algorithm). \matlab{} built-in  \\
             &      & GA    & Global optimisation method (genetic algorithm). \matlab{} built-in \\
             &      & \textbf{HGA}   & Genetic algorithm optimisation with BFGS refinement \\
\hline
\end{tabularx}

\caption{List of features of the \uqlab{} GP-module. The default values for each property is in bold.  }
\label{tbl:UQKrigingFeatures}
\end{table}

\begin{CodeInput}
KOptions.Type = 'Metamodel';
KOptions.MetaType = 'Kriging';
KOptions.ExpDesign.X = XED;
KOptions.ExpDesign.Y = YED;
myKriging = uq_createModel(KOptions);
\end{CodeInput}
The first line clarifies the type of \uqlab{} object that is being requested. Following the general UQ Framework in \figref{fig:uq_framework} a \textsc{model} object of type \code{'Metamodel'} is created. The next line specifies the type of metamodel, followed by the manual specification of the experimental design. Finally the \uqlab{} command \code{uq_createModel} is used in order to create a \textsc{model} object using the configuration options in \code{KOptions}.

The resulting Kriging metamodel object \code{myKriging} contains all the required information to compute the mean and variance of the Kriging predictor on new test points \code(X). This can be done using the following command:

\begin{CodeInput}
[meanY, varY] = uq_evalModel(myKriging, X);
\end{CodeInput}
where \code{meanY} corresponds to the mean and \code{varY} to the variance of the Kriging predictor on the test points (see Eqs.~(\ref{eq:TheoryPredicorMean}),~(\ref{eq:TheoryPredicorVariance})).

Once the metamodel is created, a report of the main properties of the Kriging surrogate model can be printed on screen by:
\begin{CodeOutput}
uq_print(myKriging);

%-------------- Kriging metamodel --------------%
Object Name:       Model 1
Input Dimension:   1

Experimental Design
  Sampling:        User
  X size:          [8x1]
  Y size:          [8x1]

Trend
  Type:             ordinary
  Degree:           0

Gaussian Process
  Corr. Type:       ellipsoidal(anisotropic)
  Corr. family:     matern-5_2
  sigma^2:          4.787983e+01
Estimation method:  Cross-Validation

Hyperparameters
  theta:	     [ 0.00100 ]
Optim. method:       Hybrid Genetic Algorithm

Leave-one-out error: 4.3698313e-01
%-----------------------------------------------%
\end{CodeOutput}

\begin{figure}[!ht]  \captionsetup{width=14cm}
	
	\begin{center}
		\includegraphics[height=8cm,keepaspectratio]{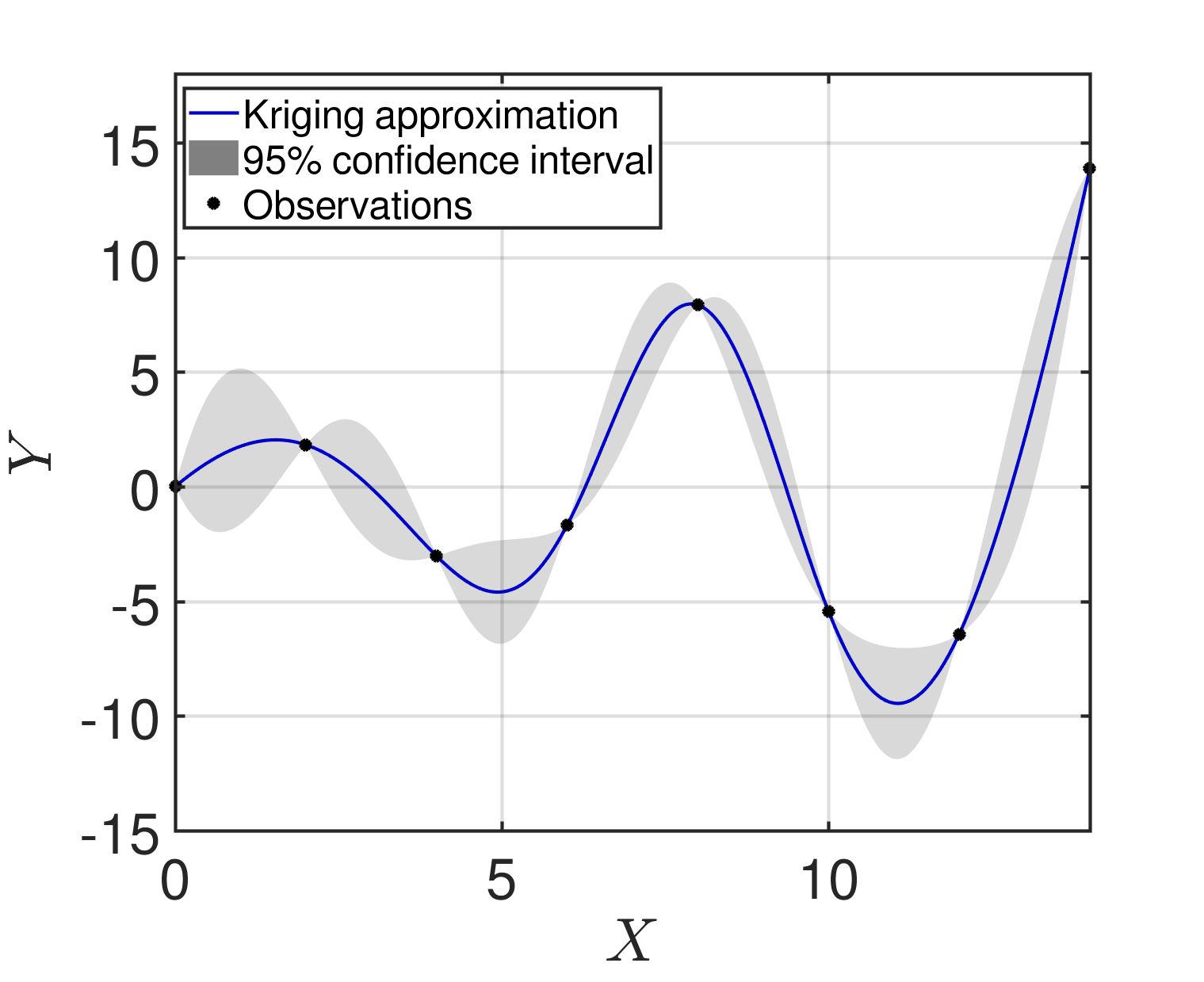}
		\caption{The output of {\code{uq\_display}} of a Kriging {\MO{}} object having a one-dimensional input.}
		\label{fig:uqdisplay}
	\end{center}
\end{figure}

It can be observed that the default values for the trend, correlation function, estimation and optimisation method have been assigned (see \tabref{tbl:UQKrigingFeatures}). A visual representation of the metamodel can be obtained by:
\begin{CodeInput}
uq_display(myKriging);     
\end{CodeInput}
Note that the \code{uq_display} command can only be used for quickly visualising Kriging surrogates when the inputs are one- or two-dimensional. The figure produced by \code{uq_display} is shown in Figure \ref{fig:uqdisplay}. 

\section{Application examples} \label{sec:Applications}

\subsection{Basic example} \label{sec:Ex1:Basic}
The goal of this introductory example is to calculate a Kriging surrogate of a well-known surrogate modelling benchmark, the Branin-Hoo function.
This function has been traditionally used as a benchmark for global optimisation methods (see \eg{}\citet{Jones1998}). A slightly modified version is considered this work, that was first proposed as a surrogate modelling benchmark by \citet{Forrester2008} due to its representative shape with respect to engineering applications. 
It is an analytical function given by:

\begin{equation} \label{eq:Ex1:BraninHoo}
	\cm(\bfx) = a \left( x_2 - b x_1^2 + c x_1 - r^2\right)^2 + s \left(1 - t \right) \cos(x_1) + s \, , \, \bfx \in \Rr^2.
\end{equation}
Some standard values of the parameters are used, namely $a=1$, $b=5.1(4\pi^2)$, $c=5/\pi$, $r = 6$, $s = 10$ and $t = 1 / (8\pi)$. The function is evaluated on the square $x_1 \in [-5, 10]$, $x_2 \in [0, 15]$.

\begin{figure}[!ht]  \captionsetup{width=14cm}
	\includegraphics[height=5.2cm,keepaspectratio=true,trim={0 0 3.5cm 0},clip=true]{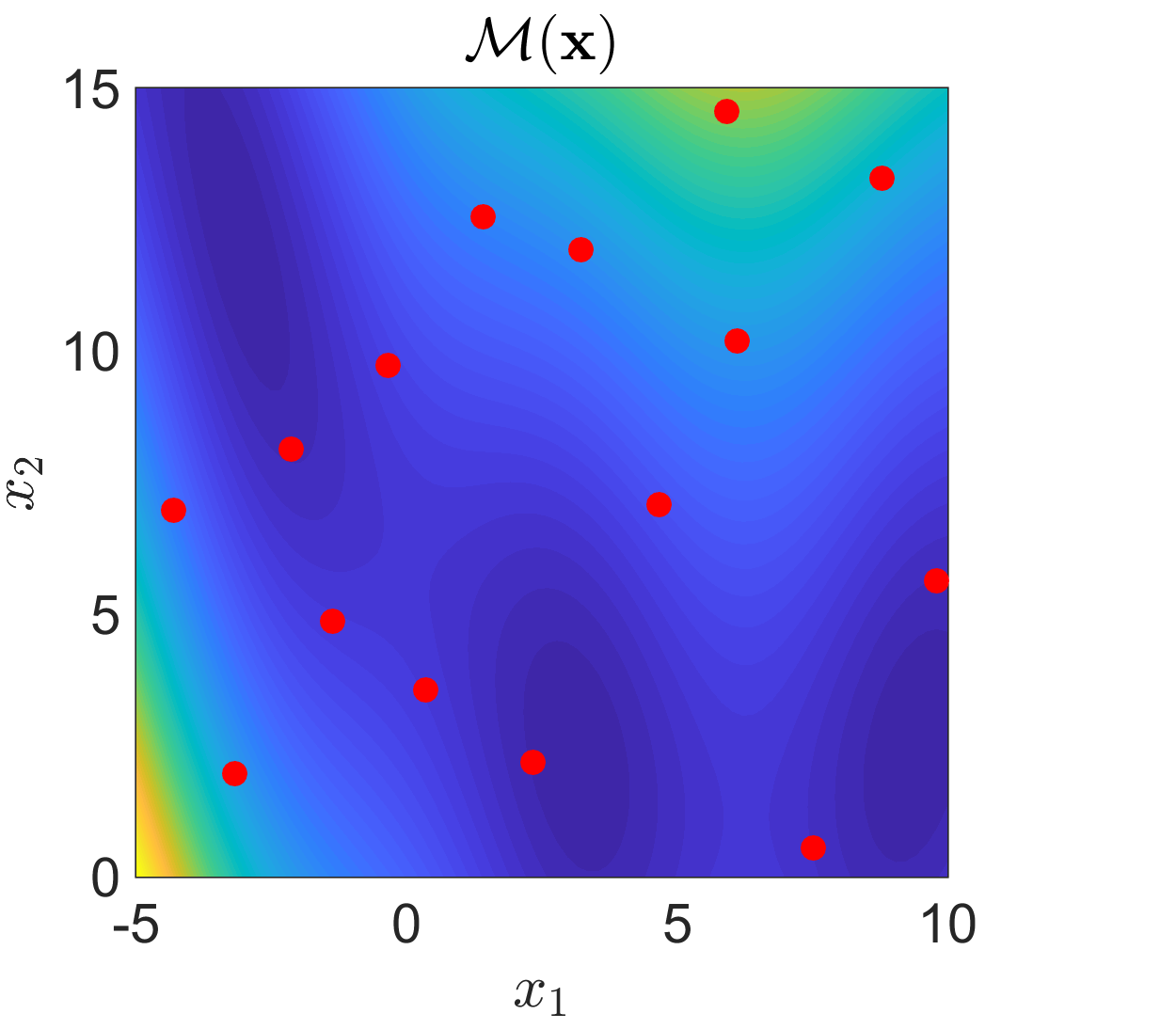}
	\includegraphics[height=5.2cm,keepaspectratio=true,trim={0.6cm 0 2.9cm 0},clip=true]{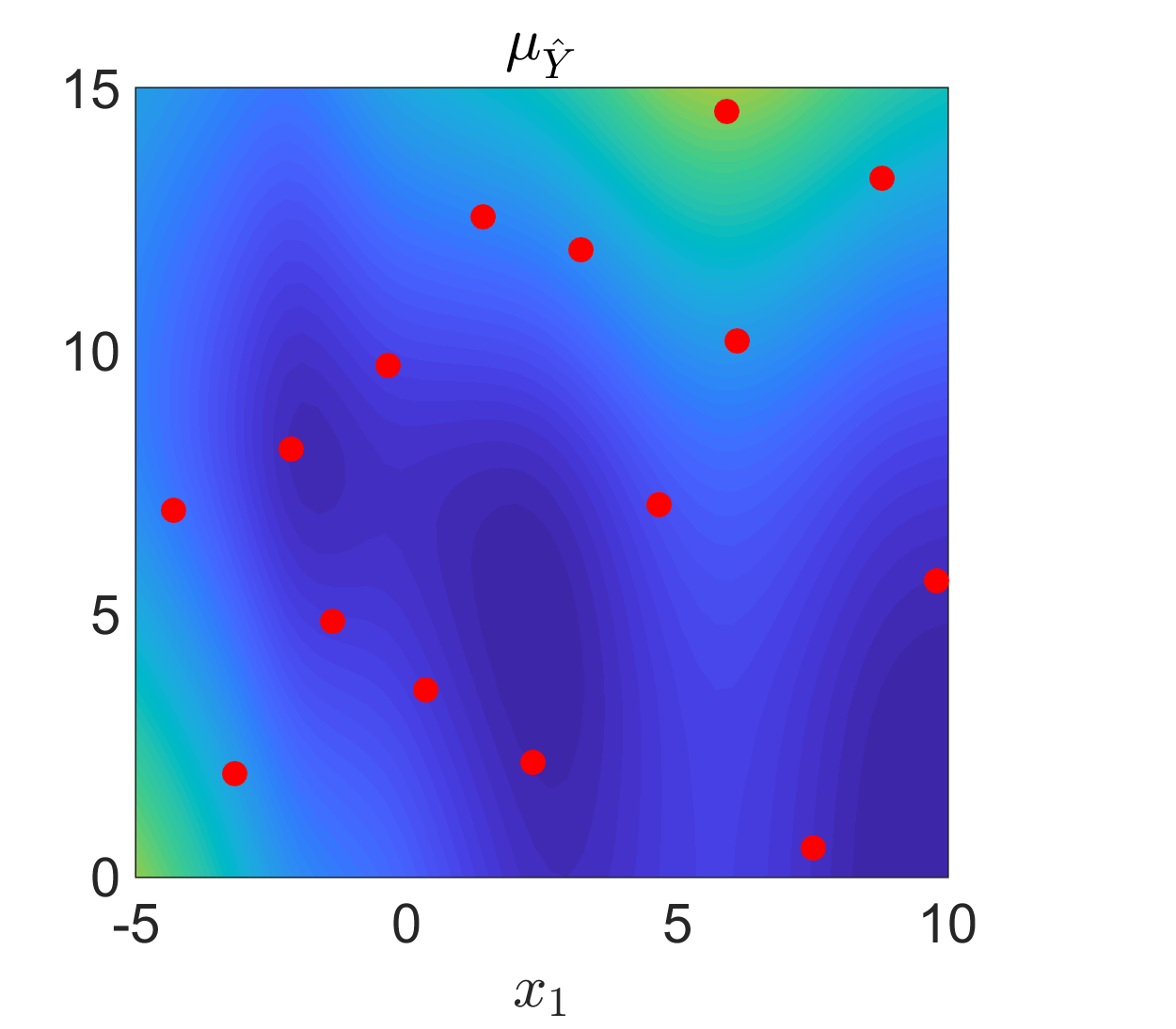}
    \includegraphics[height=5.2cm,keepaspectratio=true,trim={1.2cm 0 0 0},clip=true]{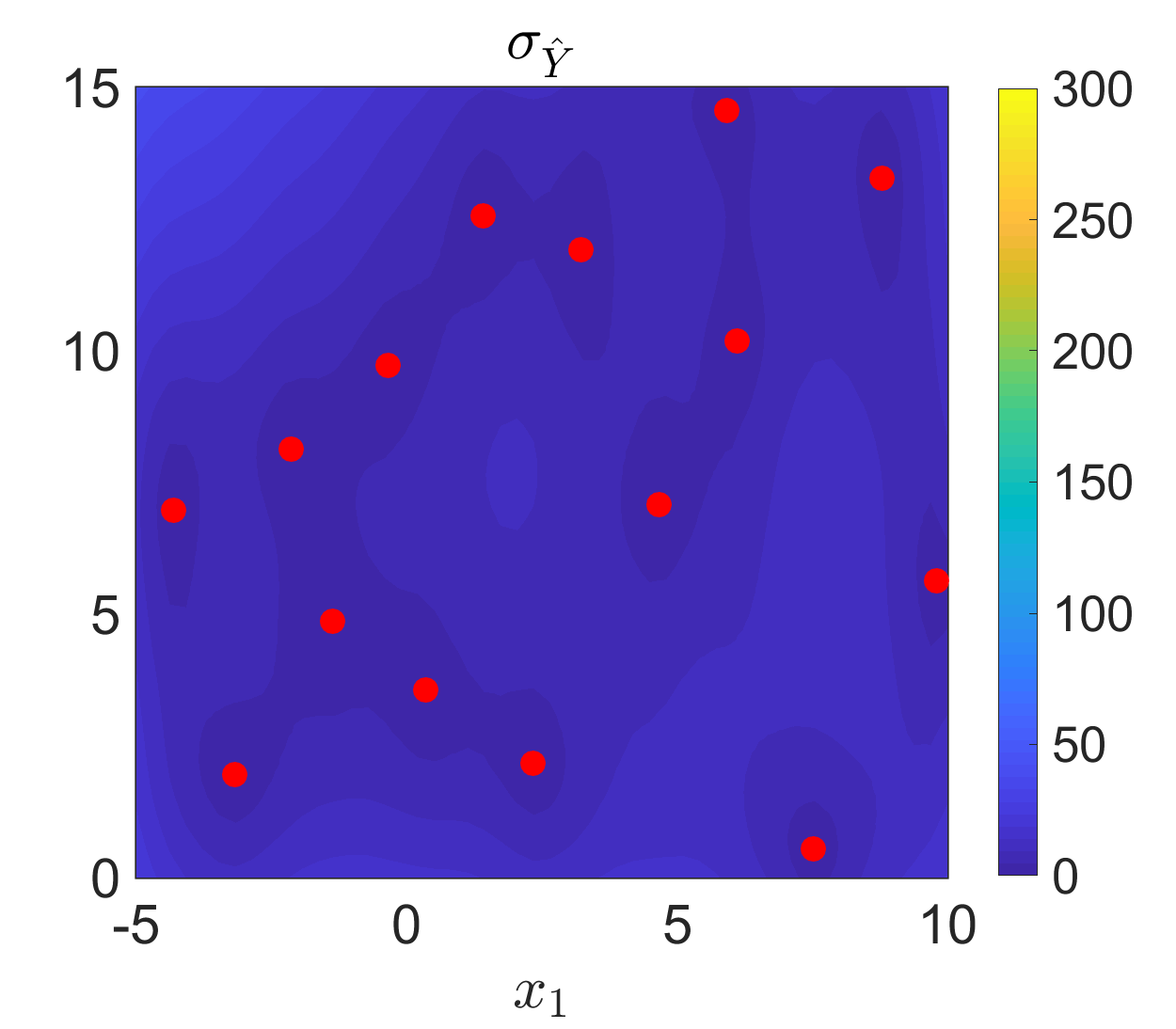}
 	\caption{From left to right: the Branin-Hoo function (true model) followed by the mean and standard deviation of the Kriging predictor. The experimental design is illustrated by red dots.}
\label{fig:Ex1:Branin}
\end{figure}

By taking advantage of the \textsc{input} and \textsc{model} modules of \uqlab{}, the experimental design and model responses that will be used for calculating the surrogate can be generated with minimal effort. First, the probabilistic input model and the true model are defined as follows:

\begin{CodeInput}
% Start the UQLab framework
uqlab;
% Specify the probabilistic input model
IOptions.Marginals(1).Type = 'Uniform';
IOptions.Marginals(1).Parameters = [-5, 10];
IOptions.Marginals(2).Type = 'Uniform';
IOptions.Marginals(2).Parameters = [0, 15];
myInput = uq_createInput(IOptions);
% Specify the computational model
MOptions.mString = ['(X(:,2) - 5.1/(2*pi)^2*X(:,1).^2 + 5/pi*X(:,1) - 6).^2' ...
'+ 10*(1-1/(8*pi))*cos(X(:,1)) + 10'];
myModel = uq_createModel(MOptions);
\end{CodeInput}

Note that the \MO{} object of the Branin-Hoo function can be equally coded in a \matlab{} m-file or written as a string (which is a useful feature for simple demo functions only). 

Next, the experimental design \code{XED} is generated along with the corresponding true model responses \code{YED}. The Latin Hypercube Sampling (LHS) method is used to obtain a space-filling experimental design of $15$ samples \citep{McKay1979}:

\begin{CodeInput}
% Draw 15 samples using Latin Hypercube Sampling
XED = uq_getSample(15, 'LHS');
% Calculate the corresponding model responses        
YED = uq_evalModel(myModel, XED);
\end{CodeInput}

A Kriging surrogate model using the \code{XED}, \code{YED} variables can be created as follows:
\begin{CodeInput}
KOptions.Type = 'Metamodel';
KOptions.MetaType = 'Kriging';         
KOptions.ExpDesign.Sampling = 'user';
KOptions.ExpDesign.X = XED;
KOptions.ExpDesign.Y = YED;
myKriging = uq_createModel(KOptions);     
\end{CodeInput}

All the required ingredients for obtaining a Kriging surrogate are assigned default values unless specified by the user (see \secref{sec:UQLabKriging}). The surrogate that is obtained can be visually inspected by issuing the command:
\begin{CodeInput}
uq_display(myKriging);
\end{CodeInput}

The result of the \code{uq_display} command is shown in \figref{fig:Ex1:Branin}. The Kriging surrogate \code{myKriging} can be used like any other model (\eg \code{myModel}) to calculate its response given a new sample of the input \code{X} using the \code{uq_evalModel} function. For example, the mean predictor, \code{meanY}, of $100$ samples generated by Monte Carlo sampling can be computed as follows:

\begin{CodeInput}
X = uq_getSample(100);
meanY = uq_evalModel(myKriging, X); 
\end{CodeInput}

More information can be extracted from the Kriging predictor using a slightly different syntax. The following code:

\begin{CodeInput}
[meanY, varY, covY] = uq_evalModel(myKriging, X); 
\end{CodeInput}

allows to retrieve the $100\times 1$ Kriging mean \code{meanY}, the $100\times 1$ Kriging variance \code{varY} and the $100\times 100$ full covariance matrix of the surrogate model responses \code{covY}.

\subsection{Hierarchical Kriging}\label{sec:Ex2:coKriging}

To further illustrate the flexibility that can be achieved with the use of arbitrary trend functions, a hierarchical Kriging application is showcased. Hierarchical Kriging \citep{han2012} is one Kriging extension aiming to fuse information from experimental designs related to different physical models of different fidelity. This is achieved by first calculating a Kriging surrogate using the low-fidelity observations and then using it as the trend of the high-fidelity surrogate. This approach can be extended to more fidelity levels in a similar fashion. A set of observations and model responses is used that originates from aero-servo-elastic simulations of a wind-turbine as presented in \citet{abdallah2015fusing}. Given a set of input parameters related to the wind flow, the output of interest is the maximal bending moment at the blade root of a wind turbine. 

Two types of simulators are available for estimating the maximal bending moment given the wind conditions. A low-fidelity simulator can generate estimates of the output with minimal computation time at the cost of lower accuracy. On the other hand a high-fidelity simulator can more accurately predict the maximal bending moment at a significantly higher computational cost. In this example a total of $300$ low-fidelity and $15$ high-fidelity simulations are available. First a Kriging surrogate is computed on the low-fidelity dataset that is contained in variables \code{XED_LF, YED_LF} as follows:

\begin{figure}[!ht]  \captionsetup{width=14cm}
	\begin{minipage}[t]{0.33\textwidth}
		\centering 
		\includegraphics[width=1.0\textwidth]{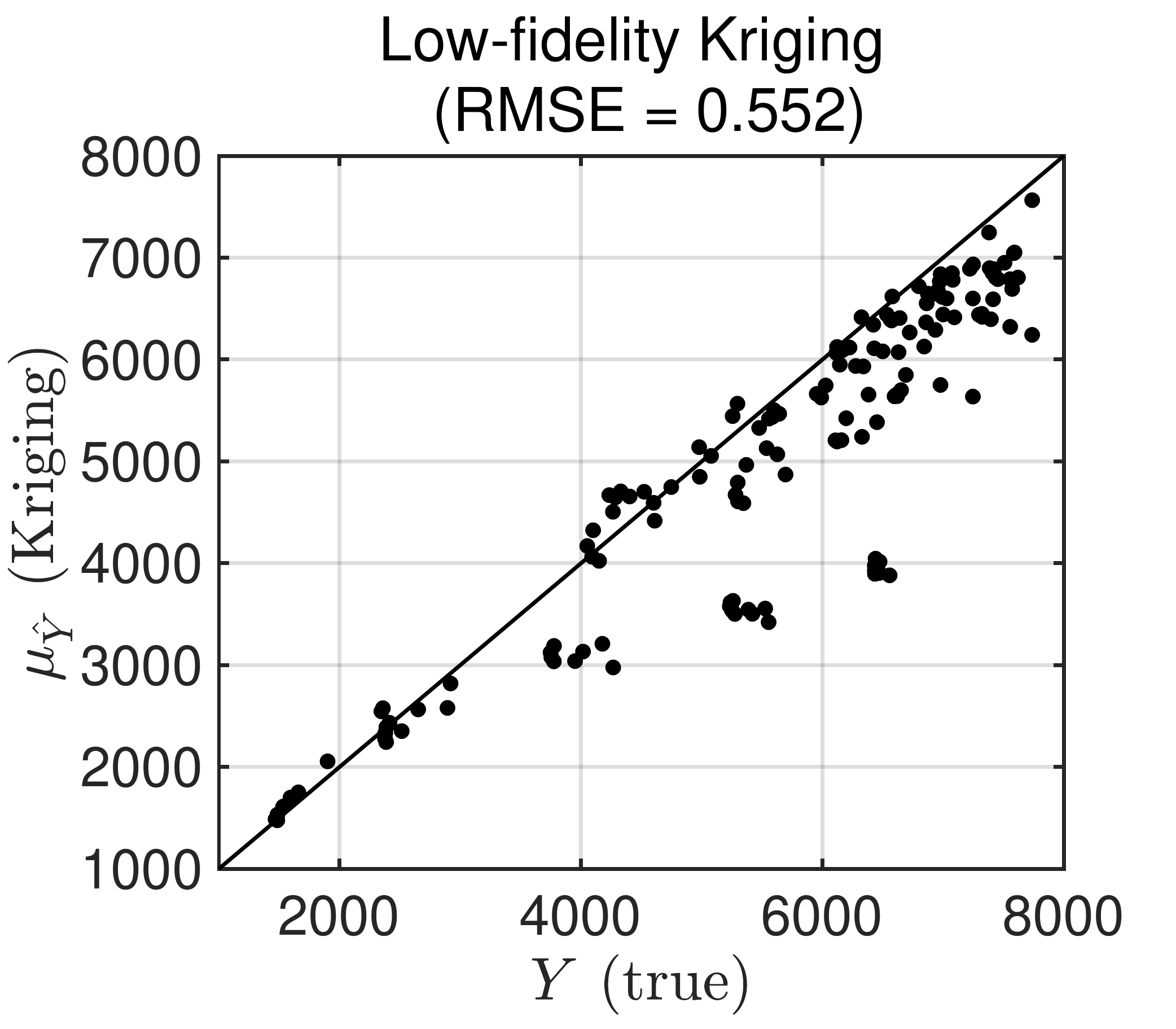}
		\label{fig:Ex2:KRG_LF}
	\end{minipage}\hfill
	\begin{minipage}[t]{0.33\textwidth}
		\centering 
		\includegraphics[width=1.0\textwidth]{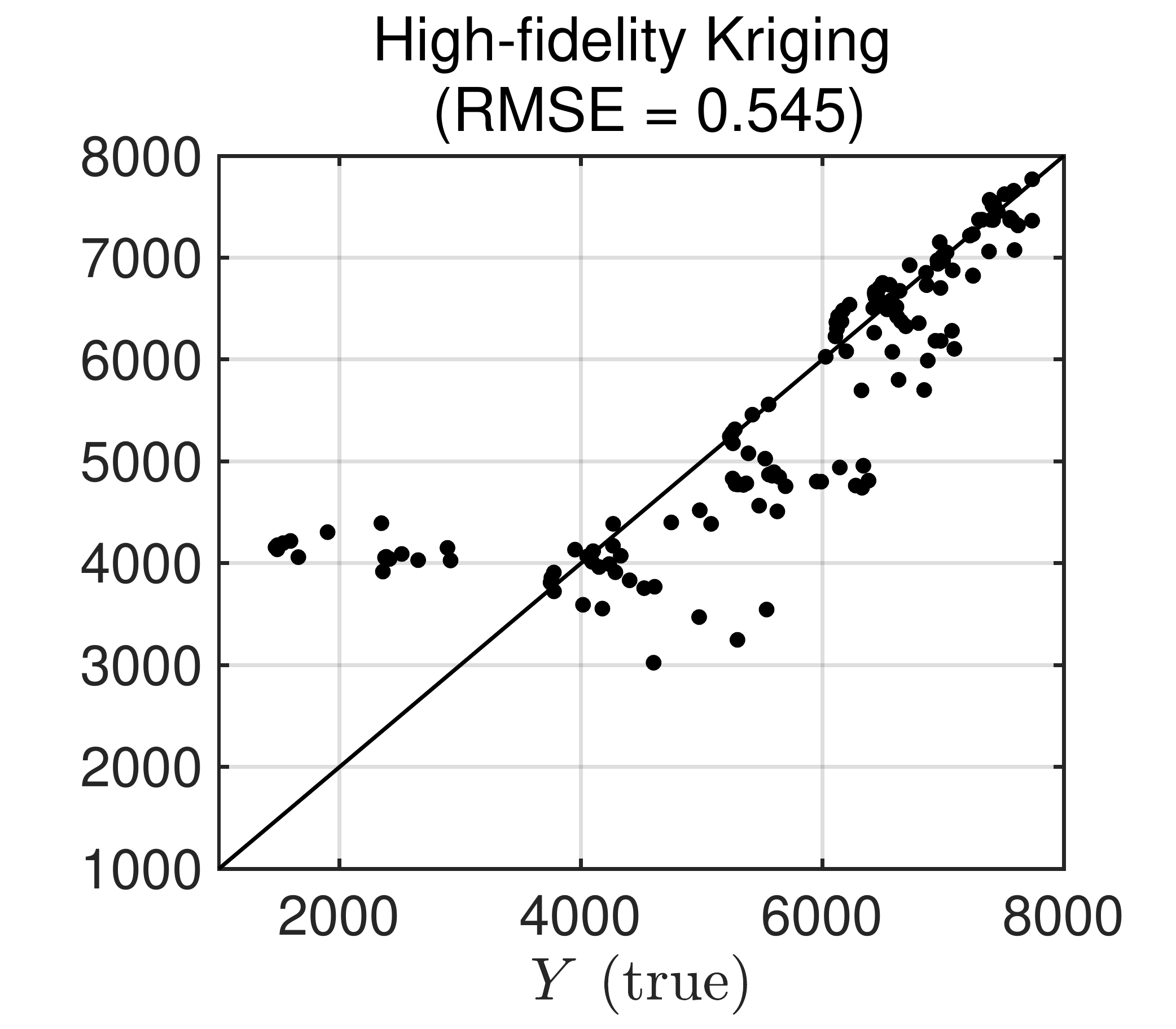}
		\label{fig:Ex2:KRG_HF}
	\end{minipage}\hfill
	\begin{minipage}[t]{0.33\textwidth}
		\centering 
		\includegraphics[width=1.0\textwidth]{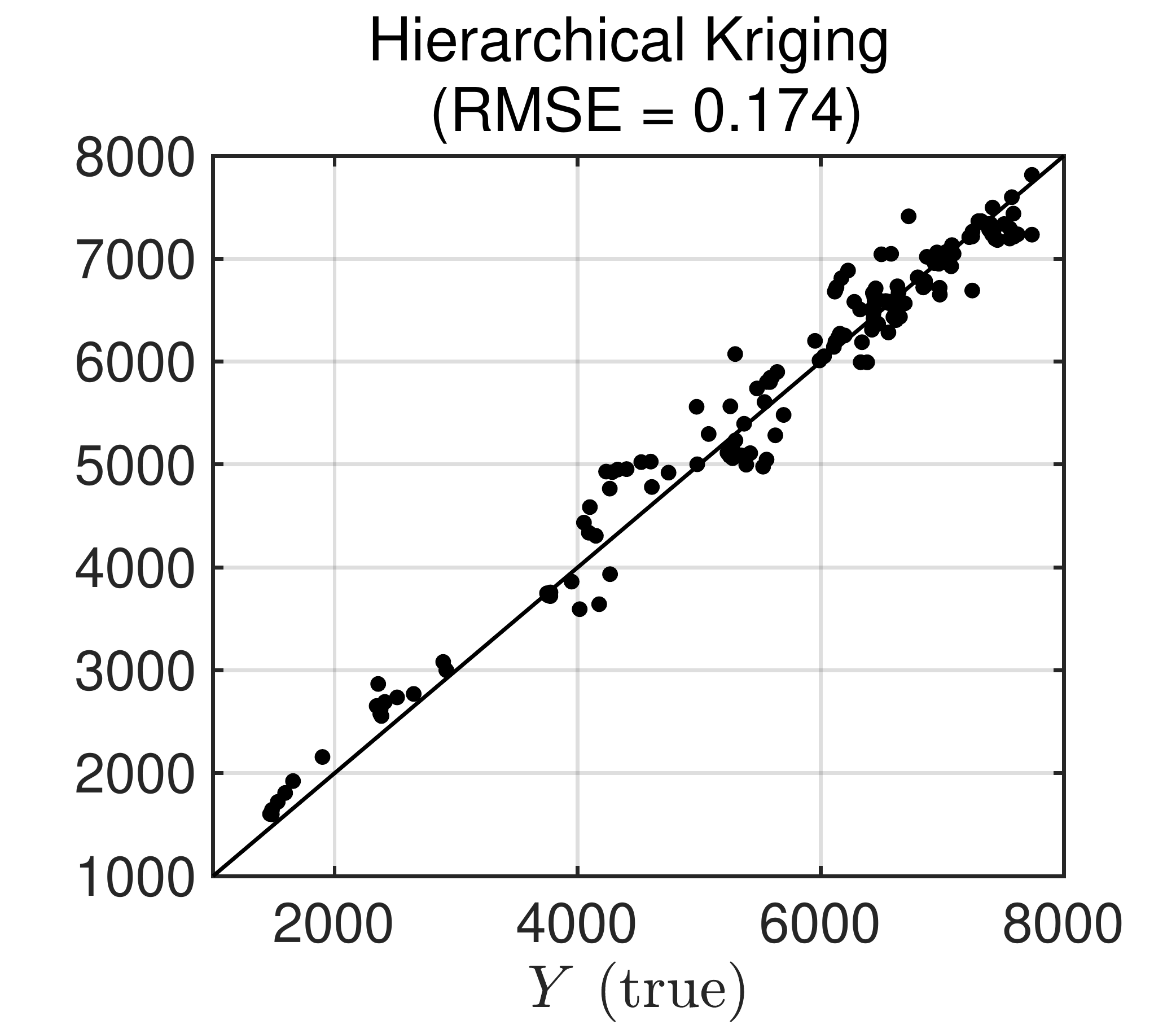}
		\label{fig:Ex2:KRG_HIER}
	\end{minipage}
	\caption{Comparison of true model output (from high fidelity simulations) versus various Kriging surrogates on a validation set of size $150$. }
	\label{fig:Ex2}
\end{figure}

\begin{CodeInput}
% Create the low-fidelity surrogate
KOptions_LF.Type = 'Metamodel';
KOptions_LF.MetaType = 'Kriging';
KOptions_LF.ExpDesign.X = XED_LF;
KOptions_LF.ExpDesign.Y = YED_LF;
KOptions_LF.Corr.Family = 'Matern-3_2';

myKriging_LF = uq_createModel(KOptions_LF);
\end{CodeInput}

Using the same configuration options, another Kriging surrogate is computed using the high-fidelity dataset (\code{XED_HF} and \code{YED_HF}):

\begin{CodeInput}
% Create the high-fidelity surrogate
KOptions_HF.Type = 'Metamodel';
KOptions_HF.MetaType = 'Kriging';
KOptions_HF.ExpDesign.X = XED_HF;
KOptions_HF.ExpDesign.Y = YED_HF;
KOptions_HF.Corr.Family = 'Matern-3_2';

myKriging_HF = uq_createModel(KOptions_HF);
\end{CodeInput}

Now a hierarchical Kriging surrogate is computed which is trained on the high-fidelity dataset but uses the low-fidelity Kriging surrogate (\ie its mean predictor) as trend:

\begin{CodeInput}
% Create the hierarchical Kriging surrogate
KOptions_Hier.Type = 'Metamodel';
KOptions_Hier.MetaType = 'Kriging';
KOptions_Hier.ExpDesign.X = XED_HF;
KOptions_Hier.ExpDesign.Y = YED_HF;
KOptions_Hier.Corr.Family = 'Matern-3_2';
KOptions_Hier.Trend.Type = 'custom';
KOptions_Hier.Trend.CustomF = @(x) uq_evalModel(myKriging_LF, x);
KOptions_Hier.Scaling = false;

myKriging_Hier = uq_createModel(KOptions_Hier);
\end{CodeInput}
The option \code{KOptions_Hier.Scaling} refers to the scaling of the input space before computing the surrogate model. In case of hierarchical Kriging scaling should be disabled because the low-fidelity surrogate is calculated on the original data and needs to be used ``as is''. 

The performance of the different surrogate models is tested on a separate validation set of $150$ high-fidelity simulations that is contained in the variables \code{XVAL_HF} and \code{YVAL_HF}. The output mean Kriging predictor on the validation set is calculated as follows:

\begin{CodeInput}
meanY_LF = uq_evalModel(myKriging_LF, XVAL_HF);
meanY_HF = uq_evalModel(myKriging_HF, XVAL_HF);
meanY_Hier = uq_evalModel(myKriging_Hier, XVAL_HF);
\end{CodeInput} 

where \code{meanY_LF}, \code{meanY_HF} and \code{meanY_Hier} correspond to the low-fidelity, high-fidelity and hierarchical Kriging predictors respectively. 

In \figref{fig:Ex2} a comparison of the true model output \code{YVAL_HF} versus the mean Kriging predictors is made. In each case the Root Mean Square Error (RMSE) is reported for quantifying the predictive performance of the surrogate:

\begin{equation}
E_{RMSE} = \frac{1}{N\Var{Y}} \sum_{i=1}^N \left( Y^{(i)} - \mu_{\widehat{Y}}^{(i)} \right)^2 
\end{equation}

where $Y$ denotes the true model outputs (in this case \code{YVAL_HF}), $\mu_{\widehat{Y}}$ the Kriging predictor mean (in this case variables \code{meanY_LF}, \code{meanY_HF} and \code{meanY_Hier} for each surrogate, respectively) and $N$ the number of samples in the validation set. 

In this example, by taking advantage of the low-cost, low-fidelity observations, the hierarchical Kriging predictor achieves a $68\%$ decrease of the RMSE on the validation set compared to the Kriging model that was solely based on the high-fidelity measurements. Moreover, by inspecting the mean responses of each Kriging predictor in \figref{fig:Ex2} it is clear that the hierarchical Kriging surrogate significantly reduces the prediction bias compared to the low- and high-fidelity ones taken as standalone. As demonstrated by this application, building a hierarchical Kriging surrogate model requires minimal effort thanks to the customisability of the GP-module.

\subsection{Kriging with custom correlation function}\label{sec:Ex3:Corr}

This example illustrates how the correlation function customisation capabilities of the GP-module  can be used to apply Kriging in a non-standard setting. 

Consider the discontinuous subsurface model given in \figref{fig:Ex3_intro}, which may represent the distribution of some soil property (e.g. porosity) in the presence of a fault. The true model consists in two realisations of two distinct random processes on the two regions $A_1$ and $A_2$ at the left and right of the fault, respectively:

\begin{figure}[!ht]  \captionsetup{width=14cm}
    \centering
    \includegraphics[height=6.5cm,keepaspectratio=true]{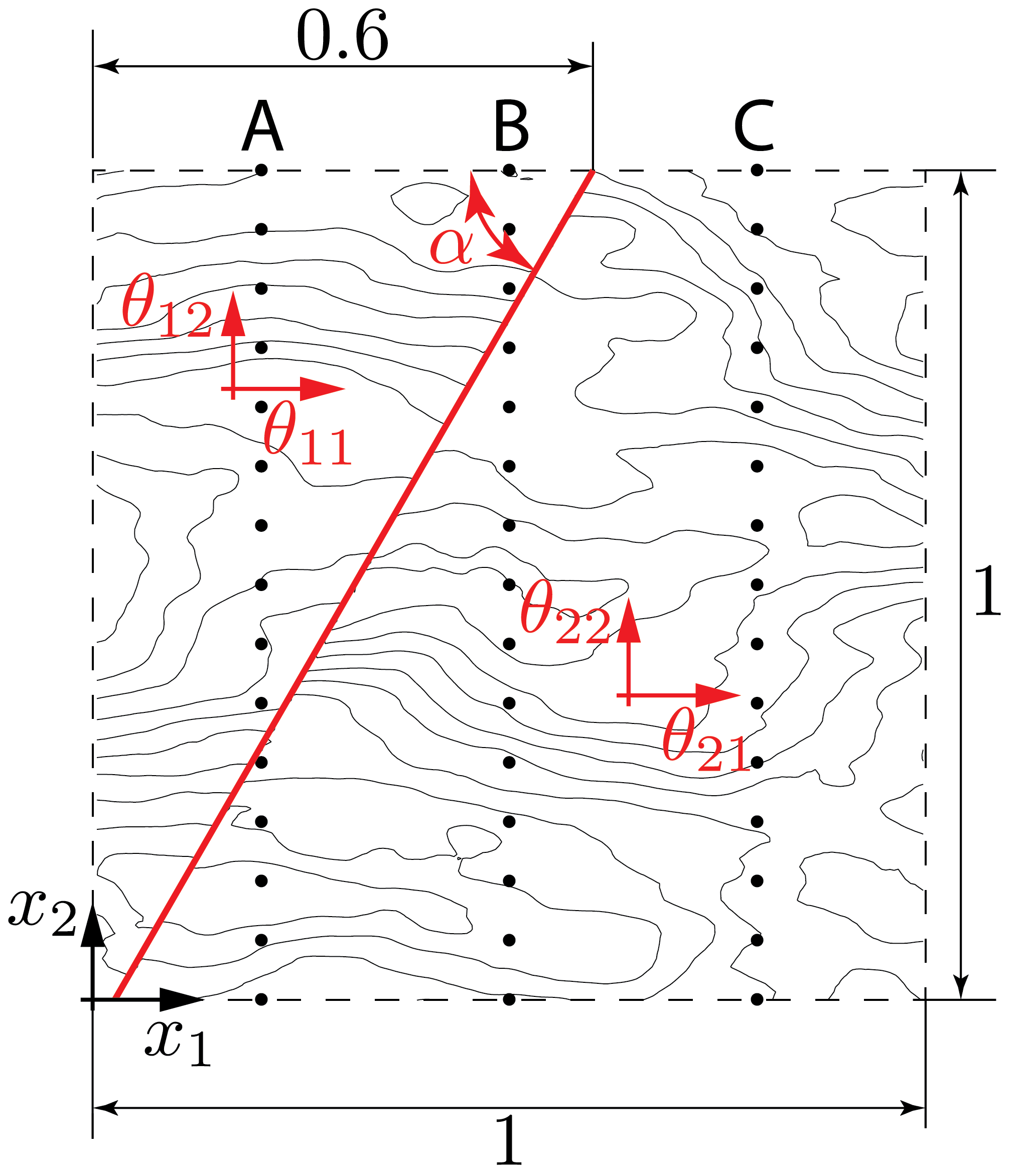}
    \caption{Graphical visualisation of the subsurface model. The unknowns (length scales of each random field and the fault angle)  are denoted by red colour.  }
    \label{fig:Ex3_intro}
\end{figure}

\begin{equation} \label{eq:Ex3_true_model}
	\cm(\bfx) = \left\{\begin{matrix}
	Z_1(\bfx, R(\bfth_1)),& \bfx \in A_1 \\ 
	Z_2(\bfx, R(\bfth_2)),& \bfx \in A_2\\ 
	\end{matrix}\right.
\end{equation}

where $\ve{x} = \acc{x_1,x_2}$ represents the spatial coordinates in the 2D domain, $Z_1$ (resp. $Z_2$) are realisations of a Gaussian process characterised by a correlation function with length scales $\bfth_1 = \acc{\theta_{11},\theta_{12}}$ (resp. $\bfth_2 = \acc{\theta_{21},\theta_{22}}$).

A Kriging surrogate model will be calculated using the following correlation function:
 
\begin{equation} \label{eq:Ex3_correlation}
R(\bfx,\bfx';\bfth) = \left\{\begin{matrix}
	R(\bfx,\bfx';\widehat{\bfth}_1), & (\bfx, \bfx') \in A_1 \times A_1 \\
	R(\bfx,\bfx';\widehat{\bfth}_2), & (\bfx, \bfx') \in A_2 \times A_2 \\
	0					   & \text{otherwise}
	\end{matrix}\right.
\end{equation}

where $\bfth = \acc{\bfth_1, \bfth_2, \alpha}$. There is a smooth dependence on $x_1, x_2$ within each region, but no correlation between points that belong to different regions. The boundary between the two regions is fully defined by the crack angle, $\alpha$, which is unknown and the fault location that is assumed to be known ($\acc{x_1, x_2} = \acc{0.6, 1}$).  The goal here is to use Kriging to interpolate the measurements taken at borehole locations A,B and C and estimate the $5$ unknown parameters $\bfth = \acc{\bfth_1, \bfth_2, \alpha}$. The correlation function of each region is the same, both in the true model and the Kriging surrogate, \ie it is assumed to be known. In particular, the correlation function is separable Mat\'ern $3/2$ (see \eqref{eq:corr_separable} and \tabref{tbl:TheoryCorrFamilies}). The maximum-likelihood method is selected for estimating $\bfth$.  Due to the complexity of the underlying optimisation problem a hybrid genetic algorithm with a relatively large population size and maximum number of generations is selected. 

 A \matlab{} implementation of the correlation function in \eqref{eq:Ex3_correlation} is given in Appendix~\ref{appendix:ex3_R}. This \matlab{} function is called \code{my_eval_R} in the following code snippet. 

The Kriging surrogate is created next, based on a limited set of observations contained in the variables \code{BoreholeLocations} and \code{BoreValues}, which contain the locations of the measurements along the boreholes and the value of the desired property, respectively.

\begin{figure}[!ht] \captionsetup{width=14cm}

	\includegraphics[height=5.2cm,keepaspectratio=true,trim={0 0 3.5cm 0},clip=true]{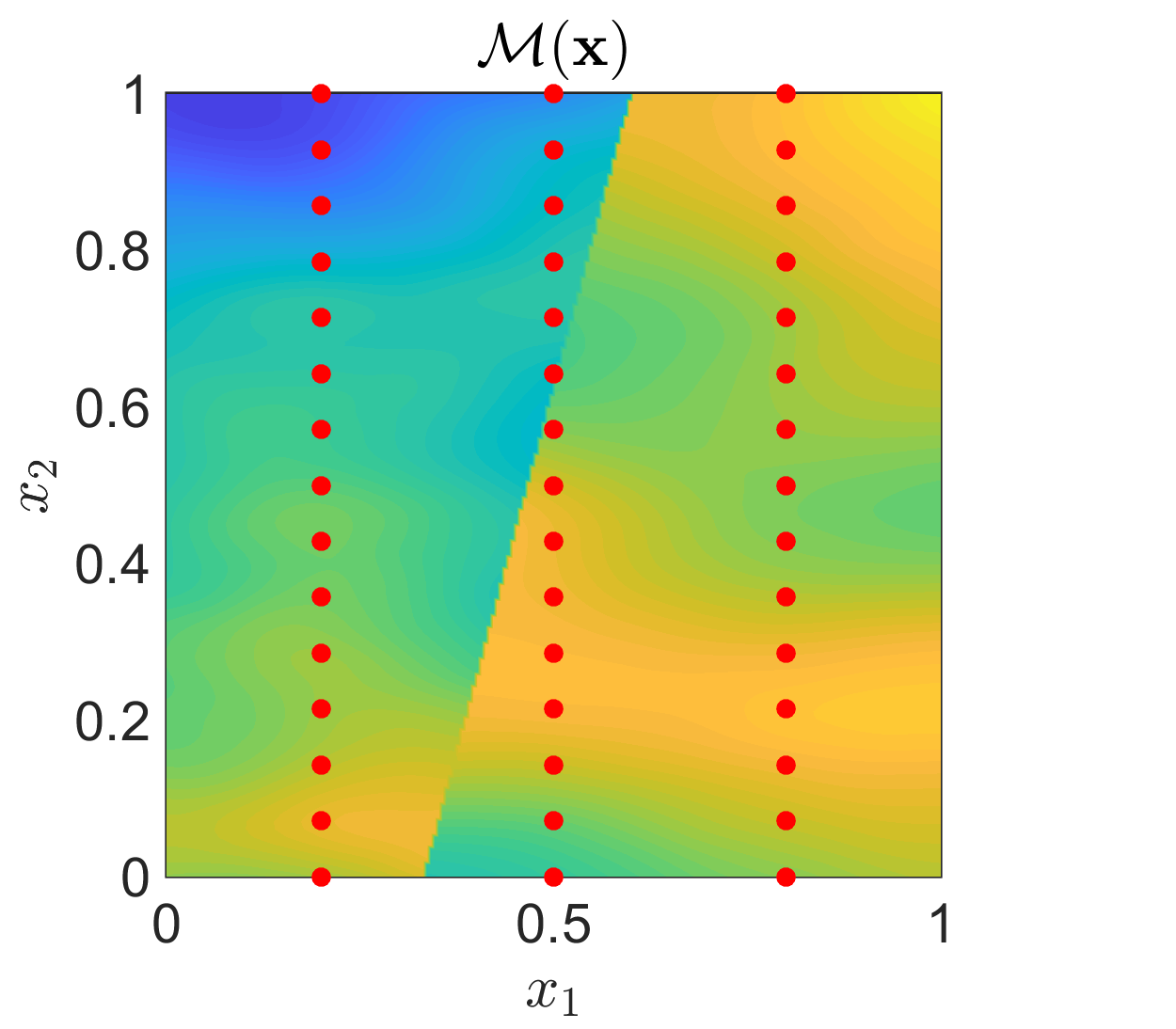}
	\includegraphics[height=5.2cm,keepaspectratio=true,trim={0.6cm 0 2.9cm 0},clip=true]{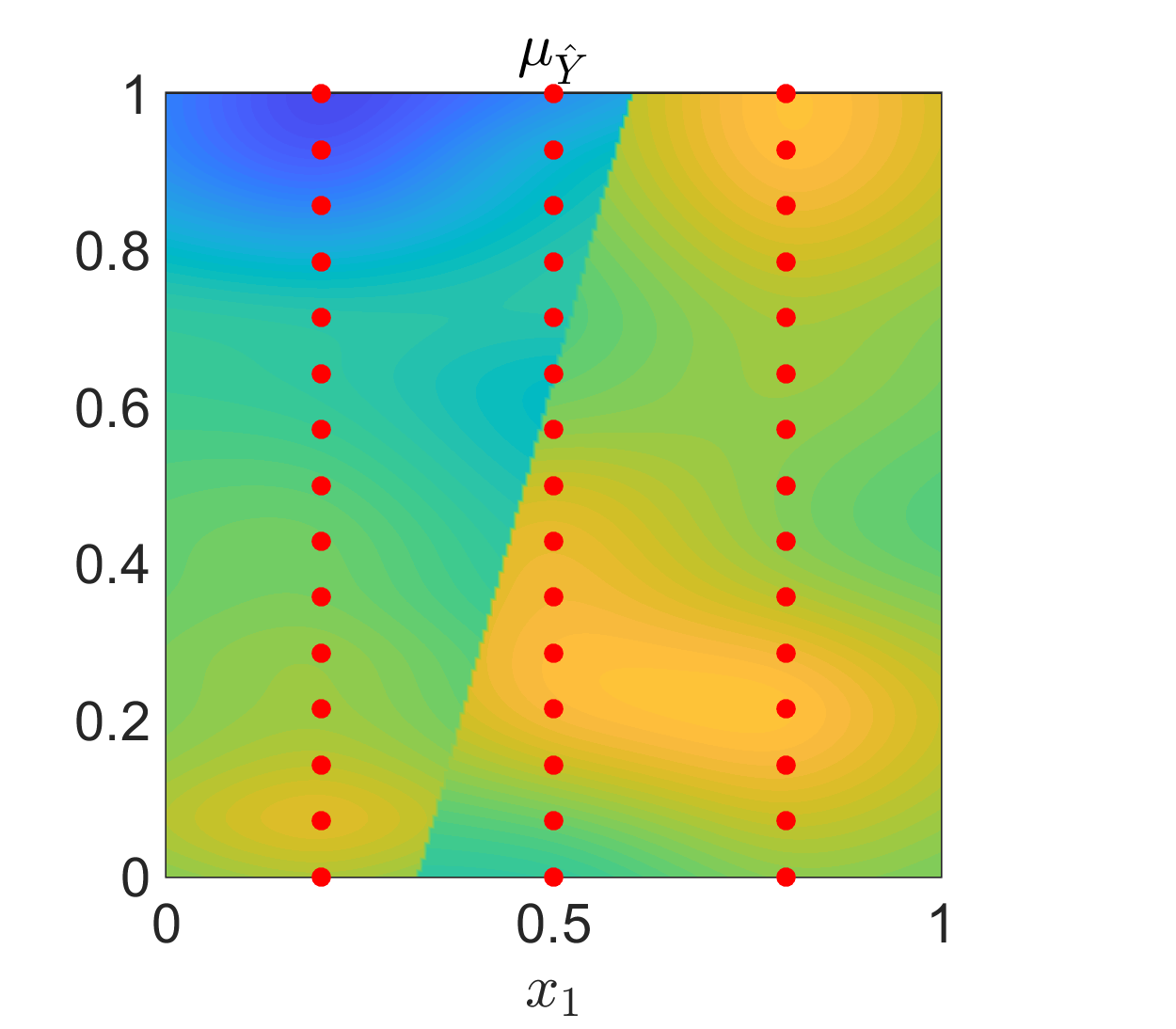}
    \includegraphics[height=5.2cm,keepaspectratio=true,trim={1.2cm 0 0 0},clip=true]{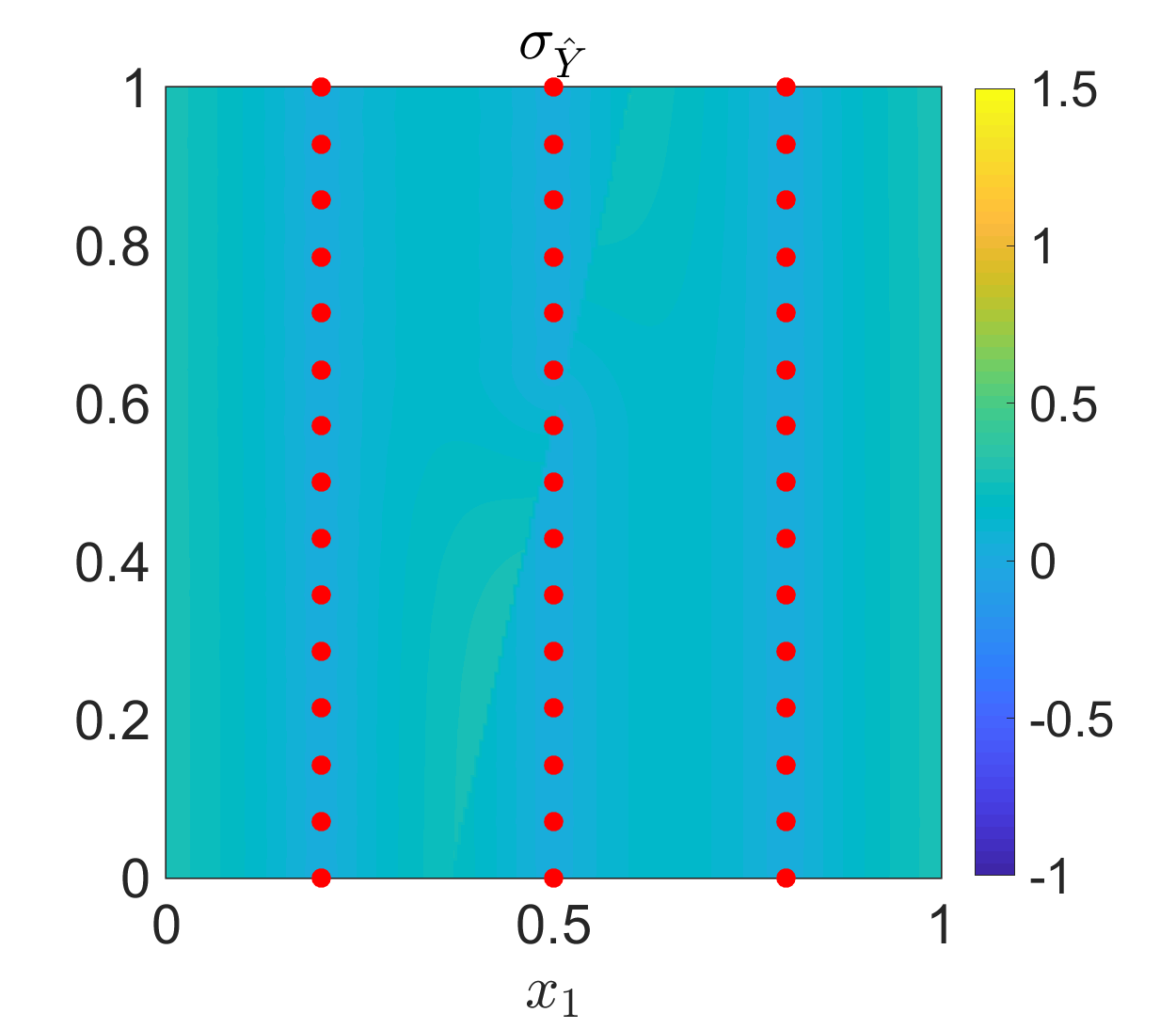}
    
	\caption{From left to right: The true permeability of the soil, followed by the mean and standard deviation of the Kriging predictor. The experimental design is illustrated by red dots.}
	\label{fig:Ex3}
\end{figure}

\begin{CodeInput}
KOptions.Type = 'Metamodel';
KOptions.MetaType = 'Kriging';
KOptions.ExpDesign.X = BoreholeLocations;
KOptions.ExpDesign.Y = BoreValues;
KOptions.Corr.Handle = @my_eval_R;
% Add upper and lower bounds on the optimization variables
BoundsL = [0.3 0.1 0.3 0.1 pi/6] ;
BoundsU = [0.9 0.5 0.9 0.5 5*pi/6] ;
KOptions.Optim.Bounds =[BoundsL ;BoundsU];
KOptions.Optim.Method = 'HGA';
KOptions.Optim.HGA.nPop = 60;
KOptions.Optim.MaxIter = 50;
KOptions.EstimMethod = 'ML';
KOptions.Scaling = False;

myKriging = uq_createModel(KOptions);
\end{CodeInput}

\begin{table}[!ht]
\centering
\begin{tabular}{l| l l l l l}
	\hline

Parameter       & $\theta_{11}$ & $\theta_{12}$ & $\theta_{21}$ & $\theta_{22}$ & $\alpha$  \\ 
\hline
True value      & $0.600$ & $0.250$ & $0.900$ & $0.350$ & $1.309$  \\ 
Estimated value & $0.310$ & $0.271$ & $0.310$ & $0.374$ & $1.342$   \\
Relative error (\%)&  $48.3$ & $8.2$ & $65.6$ & $6.9$ & $2.5$   \\
\hline
\end{tabular}

\caption{Listing of the true and estimated correlation function parameters, $\bfth$, for the Kriging surrogate of the subsurface model.}
\label{tbl:Ex3Results}

\end{table}

Once the Kriging metamodel has been computed, the mean and standard deviation of the Kriging predictor can be quickly visualised for 1D and 2D models using the \code{uq_display} command, which produces a plot similar to \figref{fig:Ex3}, except in a smaller domain determined by the range of the points in the experimental design. A comparison between the true and the estimated values of $\bfth$ is given in \tabref{tbl:Ex3Results}. As expected, the accuracy of the hyperparameters estimation is low due to the limited dispersion of the experimental design. The error of the length scale estimates along the $x_1$ direction is consistently larger due to the lack of samples along that direction.  From a coding perspective, although the correlation function that is used is relatively complex, it is straightforward to use in a Kriging surrogate once coded as a \matlab{} function (by setting the \code{KOptions.Corr.Handle} value appropriately). Moreover, custom correlation functions are allowed to have an arbitrary number of hyperparameters. The only requirement is that  the optimisation bounds (or initial value, depending on the optimisation method that is used) must have the same length as the number of the hyperparameters.  

\section{Summary and Outlook} \label{sec:Summary}

 In this paper the GP-module of the \uqlab{} software framework was presented. This \uqlab{} module enables practitioners from various disciplines to get started with Kriging metamodelling with minimal effort as was illustrated in the introductory application in \secref{sec:Ex1:Basic}. However, it is also possible to access more advanced customisation, \textit{e.g.} for research purposes. This was showcased in \secref{sec:Ex2:coKriging} where a hierarchical Kriging metamodel was developed and in \secref{sec:Ex3:Corr} where a relatively complex, non-stationary correlation function was used to solve a geostatistical inverse problem. The GP-module is freely available to the academic community since the first beta release of \uqlab{} in July $2015$.

The current version of the GP-module only allows for computing Kriging models on noisy data by explicitly providing the noise level via the nugget effect. The general case where the noise level is unknown and needs to be estimated (\textit{a.k.a.} Gaussian process regression) is currently under development and will be addressed in an upcoming release. In addition, the current version of the GP-module relies on additional \matlab{} toolboxes for performing the hyperparameter optimisation. This may be a limiting factor to some users. 

In addition to the modules currently exploiting its functionality (Polynomial Chaos-Kriging and Reliability analysis \citep{UQdoc_10_107,UQdoc_10_109}), new \uqlab{} modules that interface with the GP-module are currently under active development. The upcoming random fields module will offer several random field types (conditional and unconditional) together with advanced sampling methodologies and will be interfaced with the GP-module to offer trajectory resampling capabilities. Similarly, the upcoming Reliability-Based Design Optimisation (RBDO) module uses the surrogate modelling capabilities of the GP-module for solving RBDO
problems as described in \citet{MalikiSMO2016}.

\begin{appendix}

\section{Kriging with custom correlation function: implementation details} \label{appendix:ex3_R}

The aim of this section is to provide some additional implementation details on the application example in \secref{sec:Ex3:Corr}, in terms of the \matlab{} code involved. The correlation function described in \eqref{eq:Ex3_correlation} can be translated to the following \matlab{} function:

\begin{CodeInput}
function R = my_eval_R( x1,x2,theta,parameters )

xc = 0.6; % the x-location of the crack on the surface
yc = 1  ; % the y-location of the crack on the surface

length_scales_1 = theta(1:2);
length_scales_2 = theta(3:4);
crack_angle     = theta(5)  ;

% find the angles of each sample of x1
angles_x1 = acos(  (xc - x1(:,1))./sqrt((x1(:,1) - xc).^2 + ...
(x1(:,2) - yc).^2 ) );
% find the indices of x1 that belong to first region
idx_x1_1 = angles_x1 <= crack_angle;
% find the indices of x1 that belong to second region
idx_x1_2 = ~idx_x1_1;
% find the angles of each sample of x2
angles_x2 = acos(  (xc - x2(:,1))./sqrt((x2(:,1) - xc).^2 + ...
(x2(:,2) - yc).^2 ) );
% find the indices of x2 that belong to first region
idx_x2_1 = angles_x2 <= crack_angle;
% find the indices of x2 that belong to second region
idx_x2_2 = ~idx_x2_1;

% set-up various correlation function options so that we can re-use the
% build-in UQLab function for evaluating R in each region
CorrOptions.Type = 'separable';
CorrOptions.Family = 'Matern-3_2';
CorrOptions.Isotropic = false;
CorrOptions.Nugget = 1e-2;

% initialize R matrix
R = zeros(size(x1,1), size(x2,1));

% Compute the R values in region 1
R(idx_x1_1,idx_x2_1) = uq_Kriging_eval_R( x1(idx_x1_1,:), x2(idx_x2_1,:), ...
 length_scales_1, CorrOptions);
% Compute the R values in region 2
R(idx_x1_2,idx_x2_2) = uq_Kriging_eval_R( x1(idx_x1_2,:), x2(idx_x2_2,:), ...
 length_scales_2, CorrOptions);

end
\end{CodeInput}
The provided code, although vectorised, is optimised for readability and not performance. To that end, the internal function of the GP-module \code{uq_Kriging_eval_R} is used for calculating the correlation function value in each of the regions. 

\end{appendix}

\section*{Acknowledgements}

The authors would like to thank Dr. Imad Abdallah for providing the wind-turbine simulations dataset for the application example presented in \secref{sec:Ex2:coKriging}. 


\bibliographystyle{agsm}
\bibliography{bibliography}

\end{document}